\documentclass[12pt]{article}
\usepackage{graphicx}
\usepackage{amssymb}
\usepackage{amsfonts}
\usepackage{amsmath}
\usepackage{graphicx}
\usepackage{enumerate}
\usepackage[usenames,dvipsnames]{color} 
\usepackage[right,pagewise,displaymath, mathlines]{lineno}
\usepackage{epstopdf}
\usepackage{subfigure}

\newtheorem{definition}{Definition}
\newtheorem{note}{Note}[section]

\newenvironment{proof}[1][Proof]{\medskip \noindent{\it #1.} }{\ \rule{0.5em}{0.5em} \medskip}

\begin{document}
\bibliographystyle{cj}

\vspace*{0mm}

\noindent
{\Large\bf Measuring association via lack of co-monotonicity: \\ the LOC index and a problem of educational \\ assessment}

\vspace*{7mm}

\noindent
{\large  Danang Teguh Qoyyimi and Ri\v cardas Zitikis$^{\,*}$}

\vspace*{7mm}

\noindent
{\bf Abstract.} Measuring association, or the lack of it, between variables plays an important role in a variety of research areas, including education, which is of our primary interest in this paper. Given, for example, student marks on several study subjects, we may for a number of reasons be interested in measuring the lack of co-monotonicity (LOC) between the marks, which rarely follow monotone, let alone linear, patterns. For this purpose, in this paper we explore a novel approach based on a LOC index, which is related to, yet substantially different from, Eckhard Liebscher's recently suggested coefficient of monotonically increasing dependence. To illustrate the new technique, we analyze a data-set of student marks on mathematics, reading and spelling.

\bigskip


\bigskip


\vfill

\noindent
\rule{2cm}{0.5mm}

\smallskip

\noindent
\textbf{Danang Teguh Qoyyimi:} Department of Mathematics, Universitas Gadjah Mada, Yogyakarta 55281, Indonesia; and Department of Statistical and Actuarial Sciences, University of Western Ontario, London, Ontario N6A 5B7, Canada, E-mail: dqoyyimi@uwo.ca

\smallskip

\noindent
$^{*}$\textbf{Corresponding author: Ri\v cardas Zitikis:} Department of Statistical and Actuarial Sciences, University of Western Ontario, London, Ontario N6A 5B7, Canada, E-mail: zitikis@stats.uwo.ca

\newpage

\section{Introduction}
\label{sec.intro}

The mathematical simplicity and thus interpretability of the Pearson correlation coefficient have encouraged researchers to use it in a variety of areas where measuring association between variables is of interest. In many practical situations, however, we encounter problems that are poorly described by linear relationships and thus measuring association (or lack of it) using the Pearson correlation coefficient may not be prudent. A number of alternative ways have emerged in the literature, including the coefficients of Blomqvist, Gini, Kendall, and Spearman (cf., e.g., Nelsen, 2006).

Concisely, these coefficients provide different counting and aggregation rules of concordant and discordant pairs of bivariate data: two pairs $(x_i,y_i)$ and $(x_j,y_j)$ are concordant if either $x_i<x_j$ and $y_i<y_j$, or  $x_i>x_j$ and $y_i>y_j$. For detailed and illuminating discussions of these coefficients, we refer to Section 5.1 of Nelsen (2006), where they are also connected with the notion of copulas. For recent methodological and applied developments on copulas, we refer to Jaworski et al. (2010, 2013), and references therein.

The concordance notion leads immediately to the notion of comonotonicity that has deep roots in mathematics (cf. Denneberg, 1994; and references therein): Two functions $h$ and $g$ are comonotonic if and only if there are no $t_i$ and $t_j$ such that  $h(t_i)<h(t_j)$ and $g(t_i)>g(t_j)$. This notion has turned out to be particularly useful in economics, finance, and insurance. For details and references on the topic, we refer to, e.g., Dhaene et al. (2006), and references therein.

A number of indices for measuring dependence, concordance, and comonotonicity have been proposed in the literature (cf., e.g., Koch and  De Schepper, 2011; Dhaene et al., 2012, 2014; Liebscher, 2014; and references therein). All of them are concerned with different aspects of dependence but nevertheless -- as intended by the authors -- fall into a large class of concordance coefficients that possess certain `desirable' characteristics or properties (cf., e.g., Schweizer and Wolff, 1981; Scarsini, 1984; Nelsen, 2006; and references therein). In particular, among those characteristics is a symmetry (or interchangeability, permutation, etc.) condition, which in the context of the present paper is not desirable and would even be misleading, due to the very reason that explanatory and response variables are not symmetric (interchangeable). Hence, for measuring the lack of, or departure from, co-monotonicity between pairs of variables, none of the aforementioned coefficients can truly serve our purpose.

Nevertheless, Liebscher's (2014) suggestion for determining whether  co-movements of random variables follow an increasing pattern is philosophically closest to our current research, and we shall discuss the index briefly now, with an extensive discussion given only at the end of this paper, in Section \ref{sec.liebscher}, when all the required notions and notations have been introduced.  Specifically, given a pair of random variables, say $X$ and $Y$, whose cdf's we denote by $F$ and $G$, respectively, Liebscher's (2014) coefficient of \textit{monotonically increasing dependence} is
\begin{equation}
\label{eq.liebzeta}
\zeta_{X,Y}
= 1-{1\over c_{\psi}}\,\mathbb{E}\big [\psi\big(F(X)-G(Y)\big)\big ],
\end{equation}
where $c_{\psi}=2\int_0^1(1-u)\psi(u)du$ is the normalizing constant, and $\psi$ can be any non-negative and symmetric around $0$ function on the interval $[-1,1]$  such that $\psi(0)=0$. Various properties and extensions of this index have been discussed by Liebscher (2014), from which we see that, to a certain degree, the index can be used for tackling the problem of the current paper. Yet, due to a different goal set out by Liebscher (2014), his index does not truly serve our needs because it is 1) symmetric with respect to $X$ and $Y$ as we have noted earlier, and 2) based on rank scatterplots, whereas our problem relies on raw-data scatterplots, which can be considerably different from rank-based scatterplots as we shall see from graphs in Section~\ref{sec.liebscher}.

We have organized the rest of the paper as follows.  In Section \ref{sec.data} we describe a classical data-set of  Thorndike and Thorndike-Christ (2010), which is of our primary interest,  and then visualize the data using scatterplots with superimposed classical least-squares regression lines. In Section \ref{sec.smoothing} we fit curves to bivariate data using several powerful methods available in the literature, which is a precursor to our use of an index for measuring \textit{lack of co-monotonicity} (LOC). The definition and properties of the LOC index are discussed in Section \ref{sec.diss}, where we also provide a convenient computational formula for the index. In Section \ref{sec.subjects} we utilize the LOC index to analyze the data-set of Thorndike and Thorndike-Christ (2010). In Section \ref{sec.liebscher} we discuss the difference between the LOC index and that of Liebscher (2014). Section \ref{sec.conclusion} concludes the paper with a discussion and further references highlighting the importance of the topic that we research in the present paper. Some technicalities have been relegated to Appendix~\ref{appendix}.

\section{Data}
\label{sec.data}

To facilitate full transparency of our reasoning and adopted methodology, we use publicly available data of Thorndike and Thorndike-Christ (2010, pp.~24-25). The data  consist of marks of 52 sixth grade students on three study subjects: Mathematics, Reading, and Spelling. The students belonged to two classes, taught by two teachers, who administered tests on the three subjects. For each student and for each study subject, the teachers reported the number of correct answers and used them to assess each student's achievement on each of the three subjects.

\begin{figure}[h!]
\centering
	 \subfigure[Mathematics]{\includegraphics[scale=0.27]{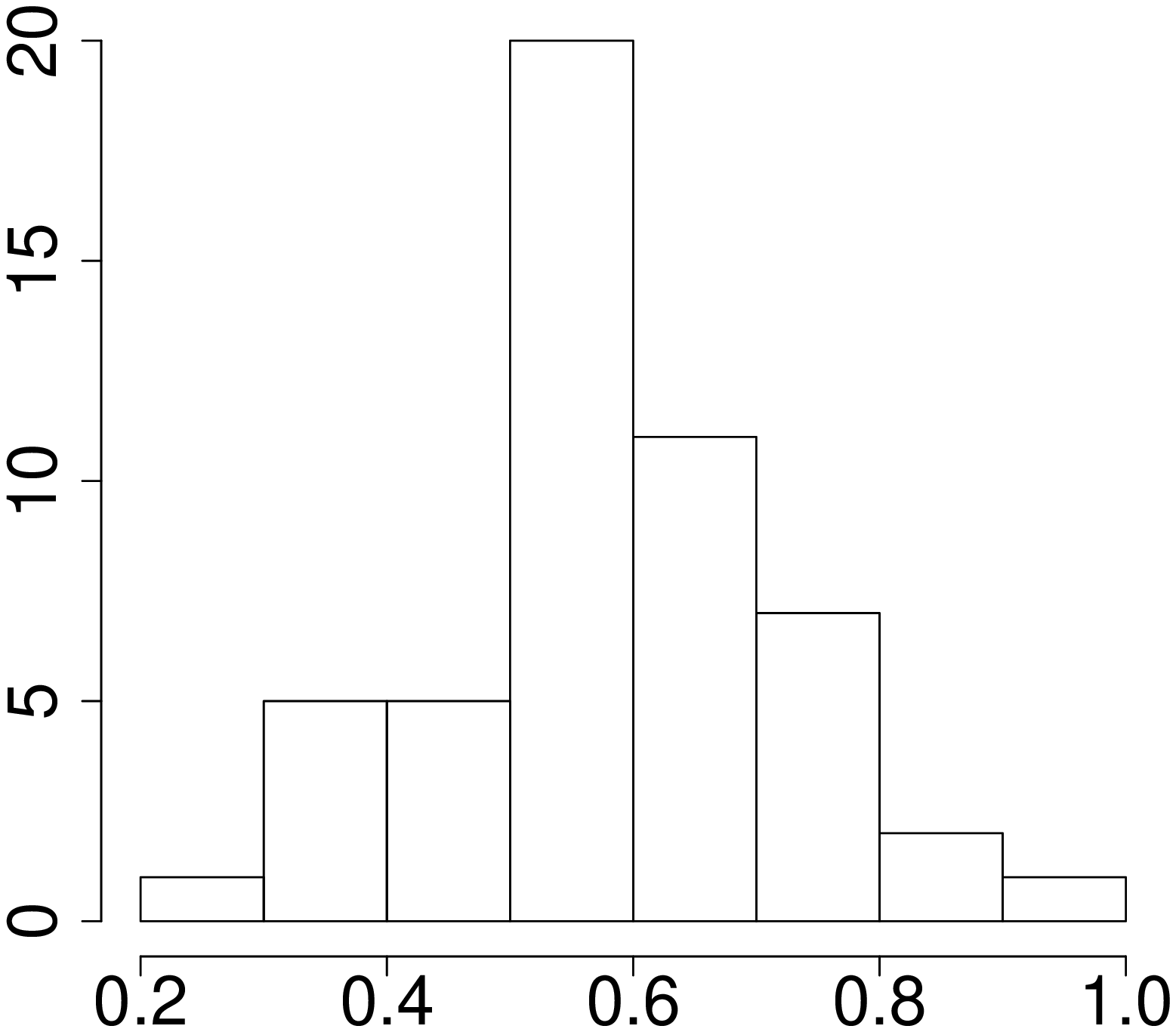}}  \quad
	 \subfigure[Reading]{\includegraphics[scale=0.27]{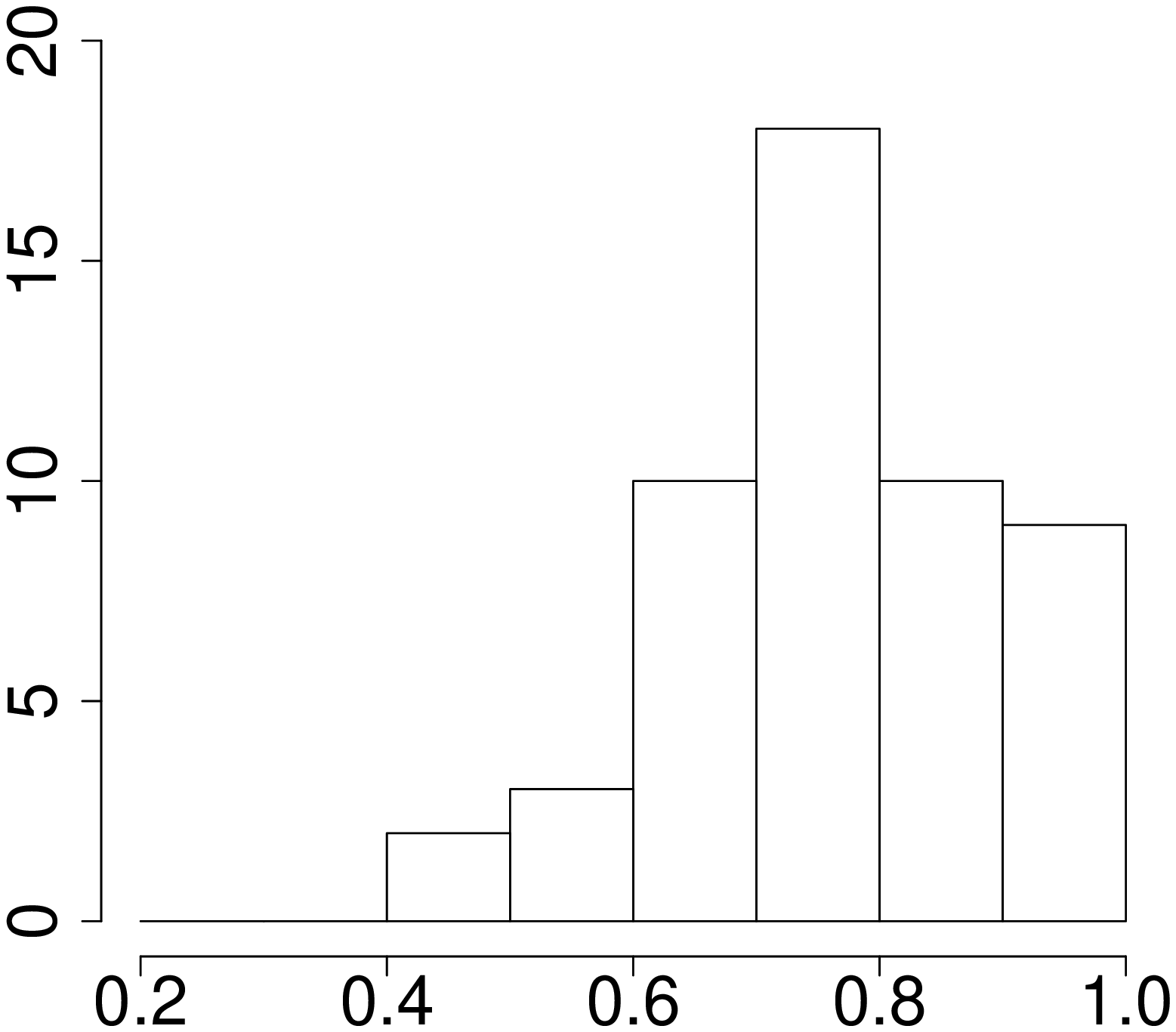}} \quad
	 \subfigure[Spelling]{\includegraphics[scale=.27]{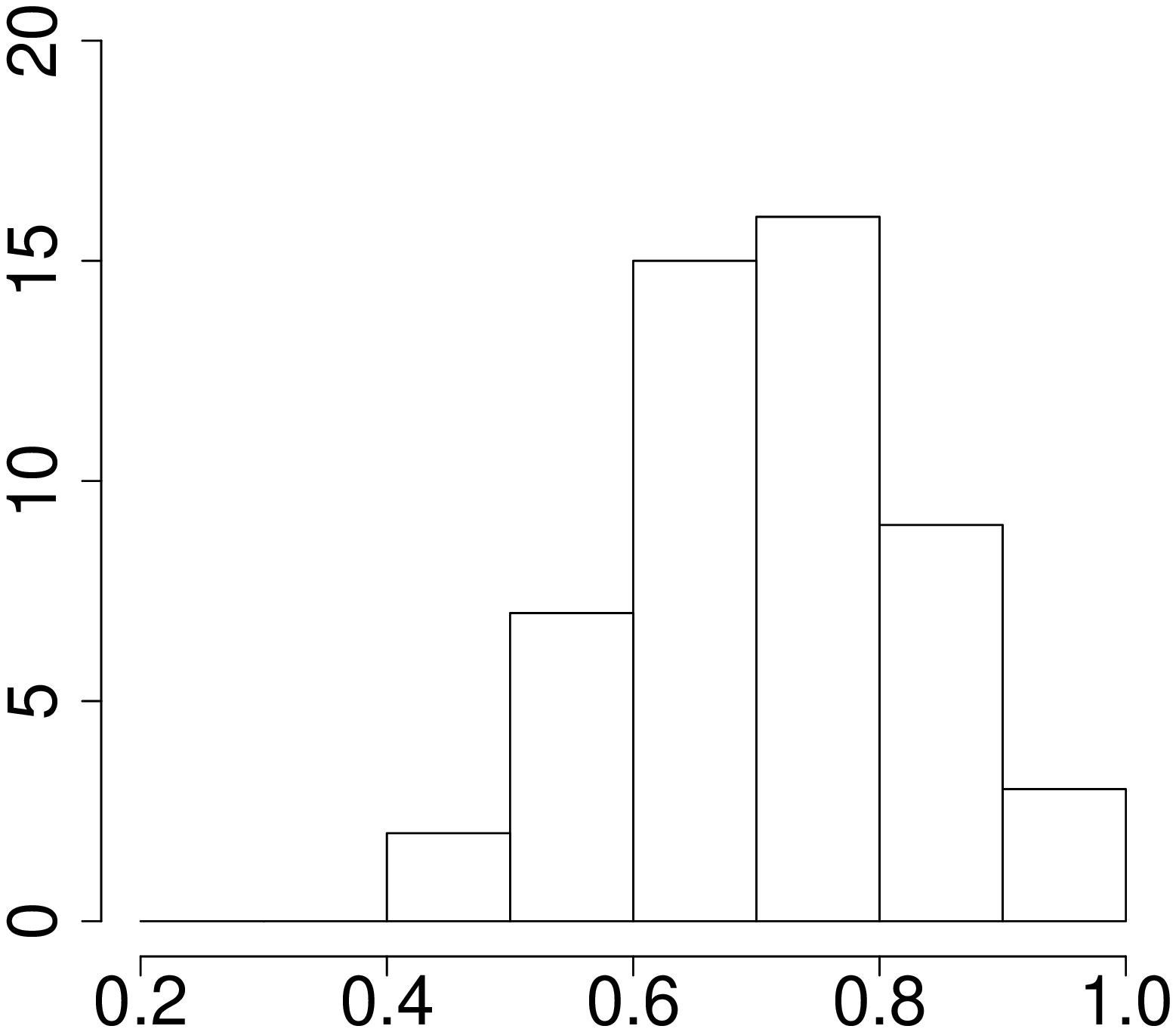}}
	\caption{Frequency histograms.}
	\label{hist}
\end{figure}

\begin{figure}[h!]
\centering
	 \subfigure[Mathematics--Reading]{\includegraphics[scale=0.25]{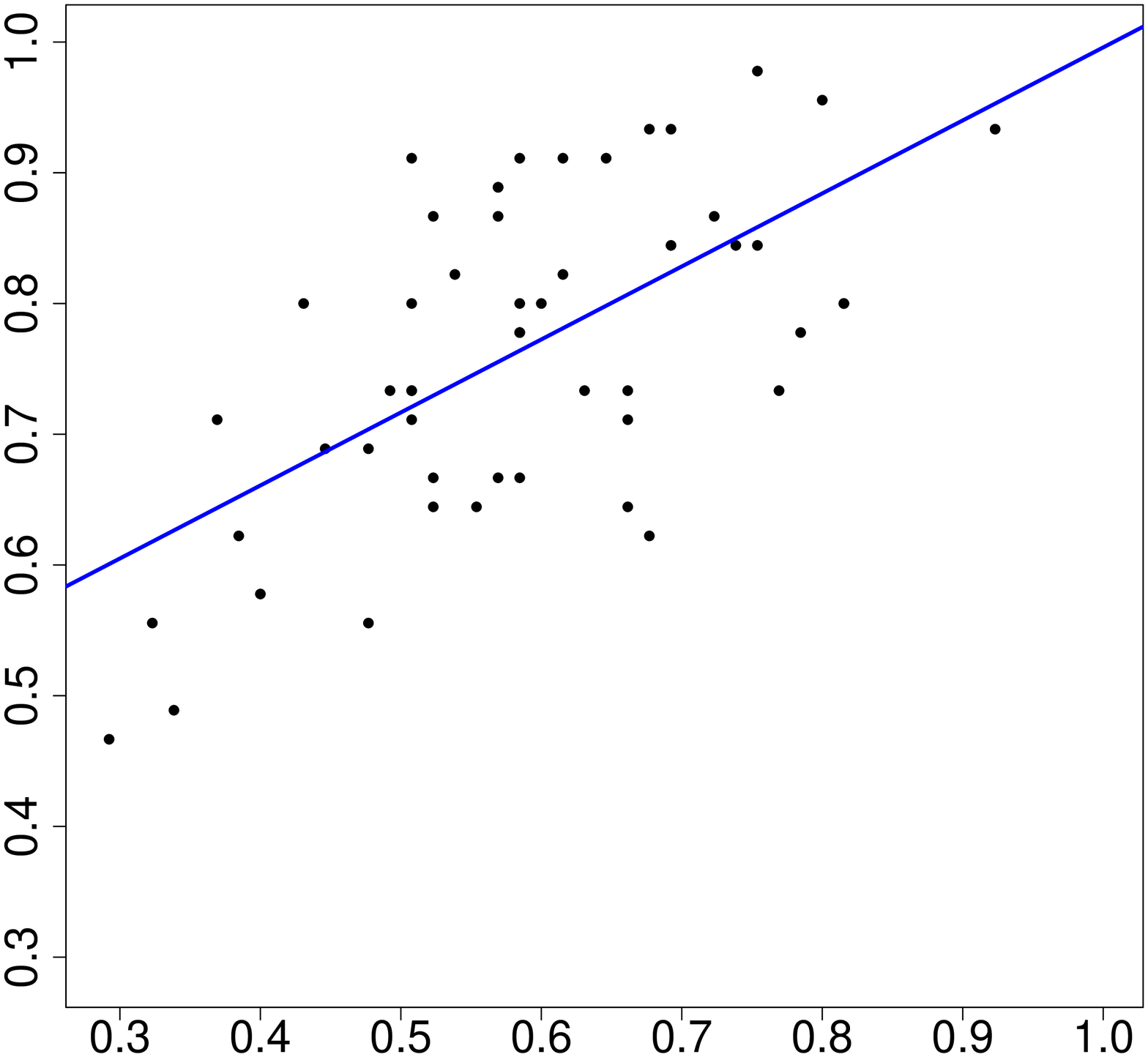}}
\qquad	 \subfigure[Mathematics--Spelling]{\includegraphics[scale=.25]{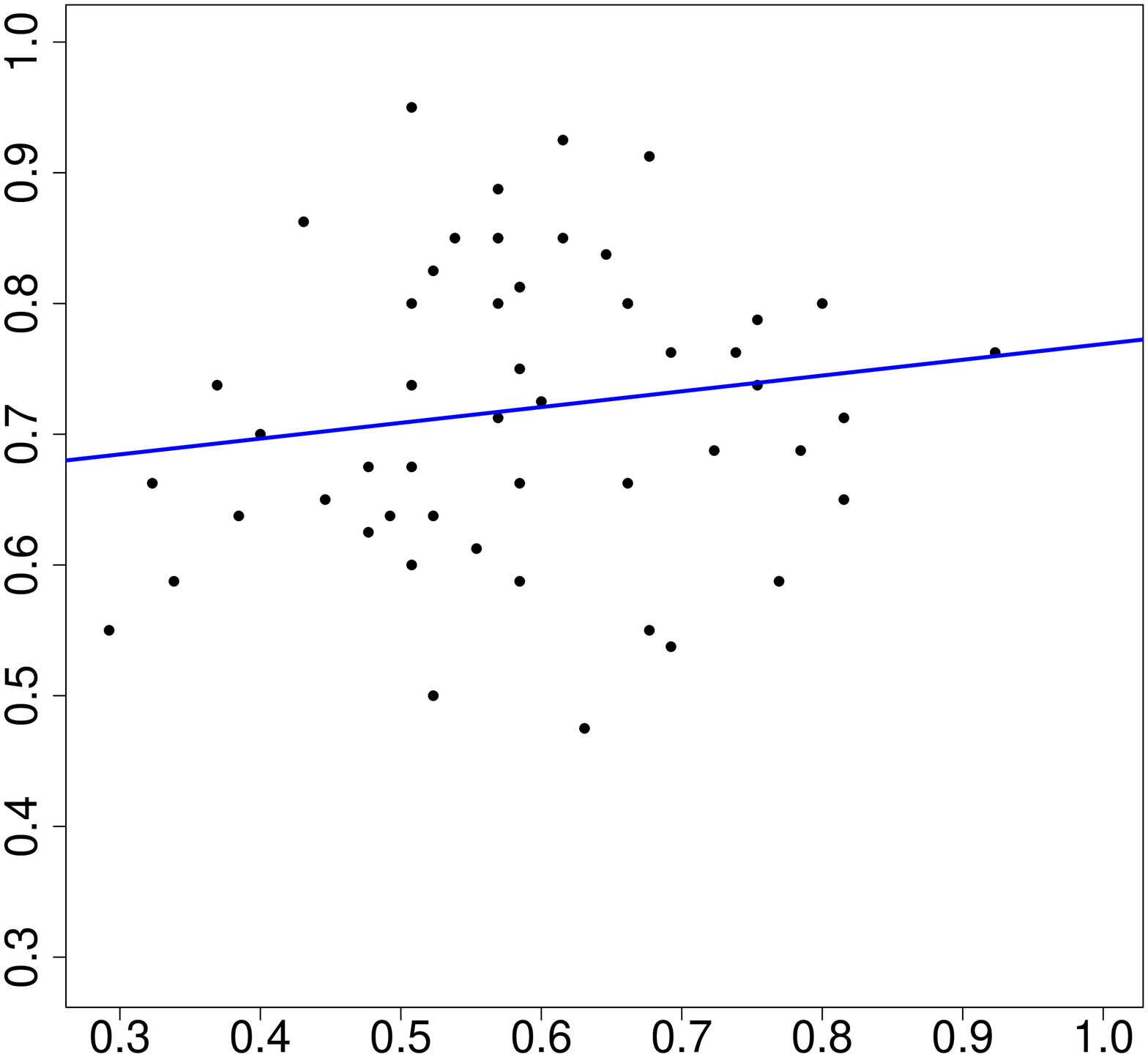}} \\
	\subfigure[Reading--Mathematics]{\includegraphics[scale=.25]{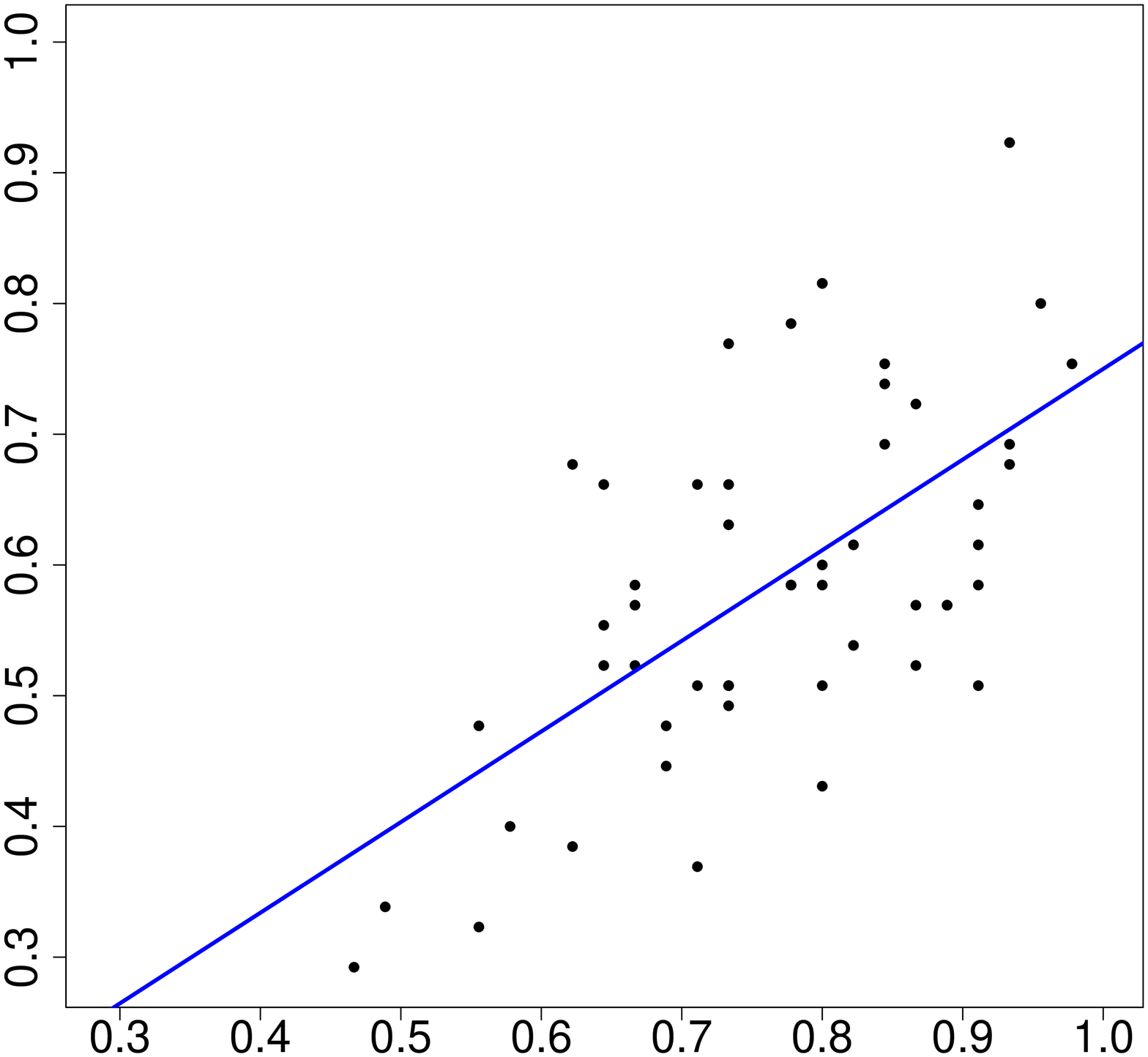}}
\qquad	\subfigure[Reading--Spelling]{\includegraphics[scale=.25]{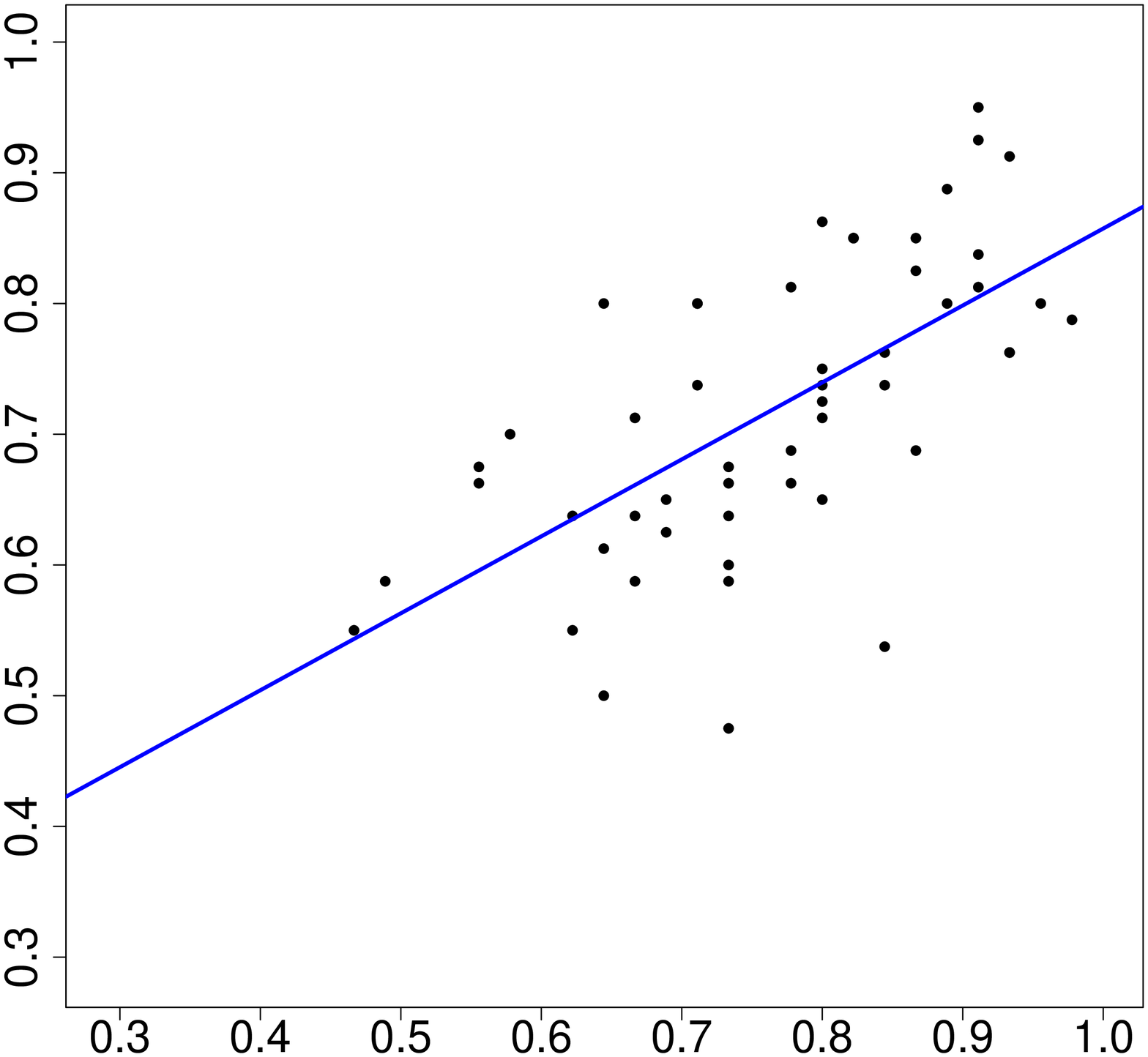}} \\
	 \subfigure[Spelling--Mathematics]{\includegraphics[scale=.25]{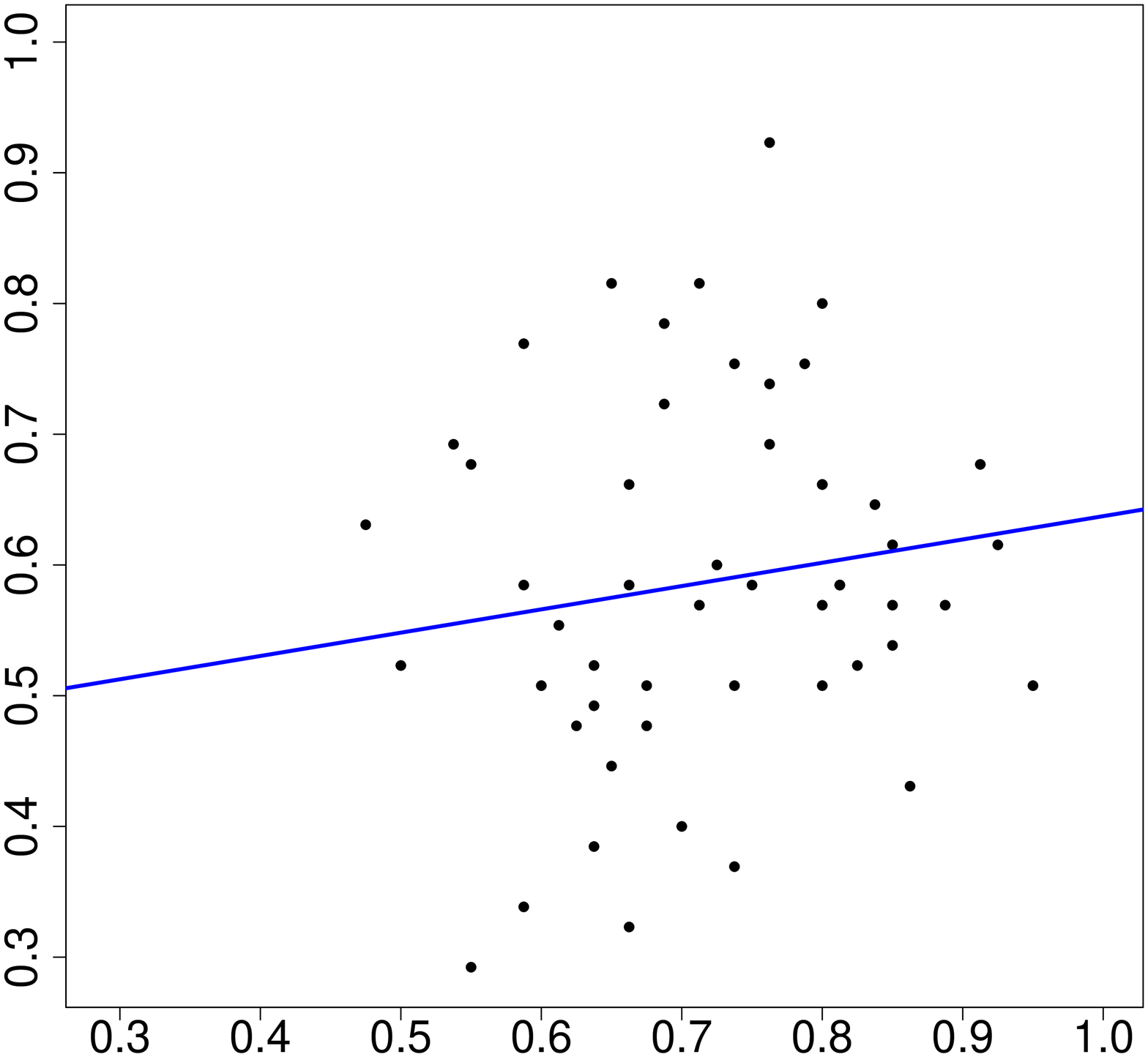}}
\qquad	\subfigure[Spelling--Reading]{\includegraphics[scale=.25]{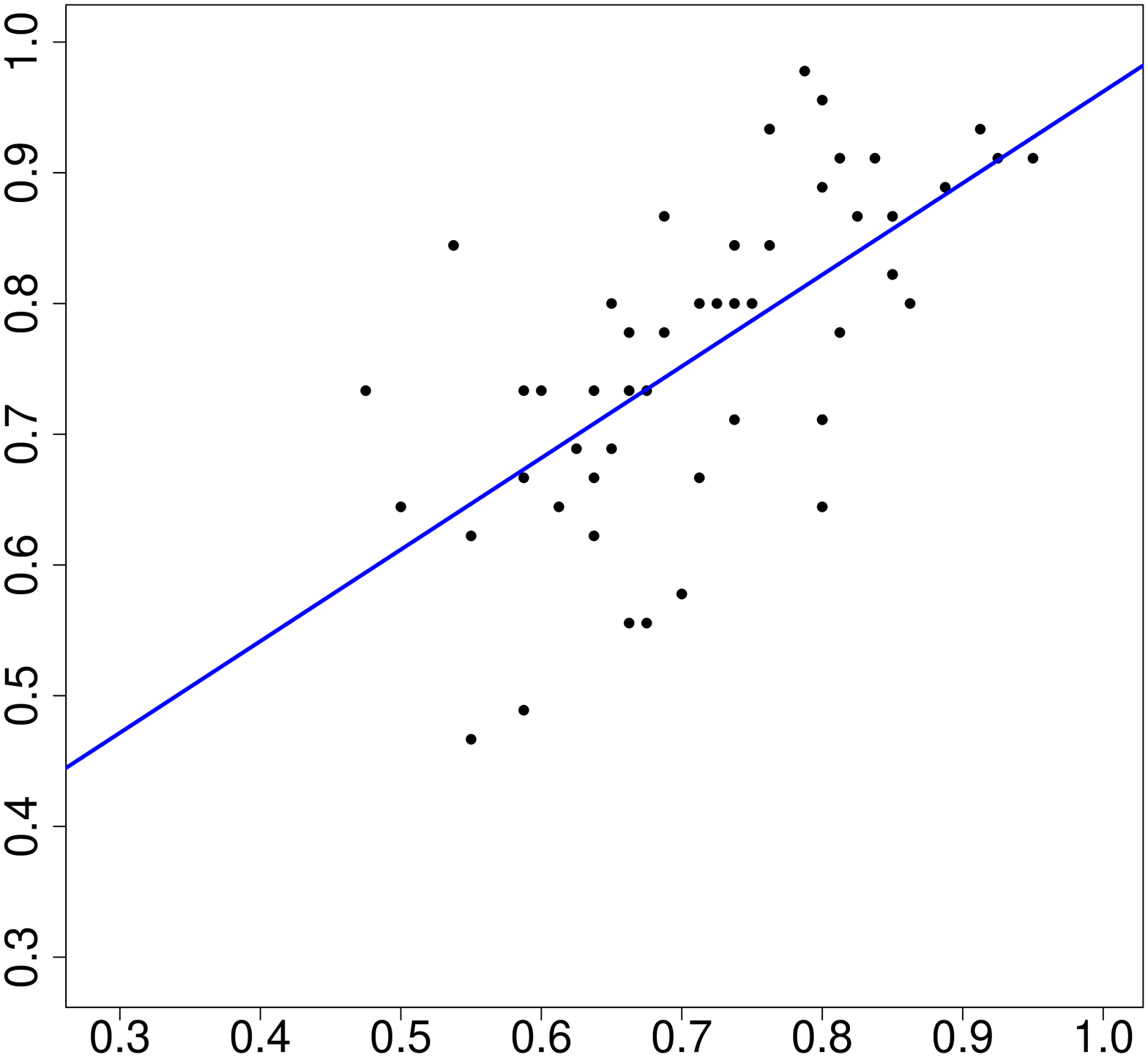}}	 
	\caption{Scatterplots and least-squares regression lines.}
	\label{los-fitting}
\end{figure}

For our analysis, we first normalize the marks to the unit interval $[0,1]$ by dividing the number of correct answers by the total number of items (i.e., questions or problems) on the tests: 65 items for Mathematics, 45 for Reading, and 80 for Spelling. Hence, throughout the paper we deal with functions $h:[0,1] \to [0,1]$ that model association between pairs of study subjects, which we denote by $X$ and $Y$, connected via the hypothetical equation $y=h(x)$ with $h$ estimated from data (topic of Section \ref{sec.smoothing}). Summary statistics and histograms of the normalized marks are reported in Table \ref{tab.descstat}
\begin{table}[h!]
	\centering
	\begin{tabular}{lccc}
		\hline
		Summary statistics  & Mathematics& Reading & Spelling   \\
		\hline
		Minimum                  & 0.2923& 0.4667 & 0.4750  \\
		1st quartile        	 & 0.5077& 0.6833 & 0.6375  \\
		2nd quartile (median) 	 & 0.5846& 0.7778 & 0.7188  \\
		3rd quartile	         & 0.6769& 0.8667 & 0.8000  \\
		Mean   	                 & 0.5873& 0.7654 & 0.7192  \\
		Maximum             	 & 0.9231& 0.9778 & 0.9500  \\
		Standard deviation		 & 0.1373& 0.1233 & 0.1129  \\
		\hline
	\end{tabular}
	\caption{Summary statistics.}
	\label{tab.descstat}
\end{table}
and Figure \ref{hist}.
In Figure \ref{los-fitting}
we have depicted the corresponding six scatterplots, which provide valuable insights into relationships between paired variables. Even though we argue that the relationships between the student marks on all pairs of the three study subjects are non-linear, it is nevertheless instructive to start our considerations with classical least-squares regression lines, which we have depicted in Figure \ref{los-fitting}, and values of the Pearson correlation coefficient, which we have reported in Table \ref{tab.corr}.
\begin{table}[h!]
\centering
\begin{tabular}{lccc}
	\hline  & Mathematics  & Reading & Spelling \\
	\hline Mathematics & 1.000000 & 0.622224 & 0.146615 \\
		   Reading     & 0.622224 & 1.000000 & 0.642215 \\
		   Spelling    & 0.146615 & 0.642215 & 1.000000 \\
	\hline
\end{tabular}
\caption{Pearson correlation coefficients.}
\label{tab.corr}
\end{table}

\section{Curve fitting}
\label{sec.smoothing}

Here we discuss curve fitting to scatterplots -- and we have six of them (see Figure \ref{los-fitting}) -- which is a precursor to calculating the LOC index, which is a topic of Section \ref{sec.diss} below.

A number of approaches have been developed for fitting curves to bivariate data. The parametric approach is one of them, which includes popular models such as linear, generalized linear, nonlinear, parametric growth curve, and many other ones (cf., e.g., Seber and Wild, 1989; and Panik, 2014). The disadvantage of this approach, especially in the context of the present paper, is that the shape and form of the functions to be fitted are difficult to guess, and thus involves an element of subjectivity that we want to avoid. Hence, we opt for the non-parametric approach, which is sometimes referred to as scatterplot smoothing (cf., e.g., Ruppert et al., 1995).

In general, there are two broad non-parametric approaches for fitting curves to bivariate data: one is based on conditional mean and another one on  conditional quantile. Both methods have their own advantages and disadvantages, and we shall illustrate both of them. We note at the outset that in the case of the conditional quantile, we shall restrict our attention to the conditional median that serves a natural alternative to the mean when data are skewed. Some further details and references on the two methods  will be provided in Section \ref{sec.hhat} below, with their actual use for analyzing the data of Thorndike and Thorndike-Christ (2010) exhibited in Section \ref{sec.subjects}.

\subsection{Constructing $\widehat{h}$}
\label{sec.hhat}

The \textit{conditional-mean} approach is based on the assumption that a good model for $h$ is given by the conditional mean, and thus
\begin{equation}
h(x) = \mathbb{E}\left[Y|X=x\right].
\label{expect}
\end{equation}
Given a scatterplot consisting of $n$ pairs $(x_i, y_i)$, the local linear estimate -- which is our choice among many other ones available in the literature -- for estimating $h(x)$ is given by
\[
\widehat{h}(x)=\widehat{\beta}_0,
\]
where $\widehat{\beta}_0$ is a solution to the minimization problem
\begin{equation}
\min_{\beta_0,\beta_1}\sum_{i=1}^{n} L(y_i-(\beta_0+\beta_1(x_i-x))K\left((x_i-x)/b\right);
\label{def.loclinrq}
\end{equation}
throughout this paper we work with the standard normal kernel $K$. Details and references on the bandwidth $b$ selection will be provided in Section \ref{sec.bandselection} below. As to the loss function $L$, in the conditional-mean case we use the quadratic loss function $L(x)=x^2$, which is a natural choice because the expected quadratic loss is minimized at the mean. In the case of the conditional-median approach, an analogous argument leads us toward the absolute loss function $L(x)=|x|$.

We note in passing that this estimate naturally arises from the fact -- recall here the local constant regression method of Nadaraya-Watson model --  that $h(x)$ defined by equation (\ref{expect}) solves the minimization problem $\mathbb{E}\left[(Y-\beta_0)^2|X=x\right]$ with respect to $\beta_0$. The additional quantity $\beta_1(x_i-x)$ in objective function (\ref{def.loclinrq}) is included to diminish the asymptotic bias of the estimate, if compared to the bias arising from the Nadaraya-Watson method (cf., e.g., Fan, 1992). For further properties of the local linear estimate, we refer to Wand and Jones (1995), and references therein.

It is also natural to use the \textit{conditional-quantile} approach (Koenker, 2005), which is based on the assumption that a good model for $h(x)$ is given by the conditional quantile, and thus
\[
h(x)=Q_{Y|X=x}(\tau)
\]
for some $\tau \in (0,1)$. An estimate $\widehat{h}(x)$ of $h(x)$ stems from  the minimization problem of (\ref{def.loclinrq}) using the loss function $L(x)$ that is equal to $\tau x$ for all $x\ge 0$ and $(1-\tau) (-x) $ for all $x<0$. Upon recalling that throughout this paper we set $\tau=0.5$, in the conditional-median case we therefore work with the absolute loss function $L(x)=0.5|x|$; the factor $0.5$ is of course irrelevant in our considerations as it does not influence the result of minimization problem (\ref{def.loclinrq}).

\subsection{Bandwidth selection}
\label{sec.bandselection}

The construction of bandwidth $b$ is based on how good the resulting estimator $\widehat{h}(x)$ of $h(x)$ is, and for this task it is customary to use the mean integrated squared error (MISE)
\begin{equation}
\mathrm{MISE}\big(\widehat{h}\big)=\int \mathbb{E}\Big [ (\widehat{h}(x)-h(x))^2 \mid x_{1}, x_{2}, \dots, x_{n} \Big ]w(x)dx
\label{def.mise}
\end{equation}
with some weight function $w$ that ensures convergence of the integral (e.g., Ruppert and Wand, 1994). Specifically, the bandwidth is chosen so that it asymptotically minimizes the MISE. There are of course other good ways to choose the bandwidth but we shall not delve deeply into this subject here and just note some of the facts that we shall utilize in our data-driven computations.

Namely, we follow Ruppert and Wand (1994), Ruppert et al. (1995), and Fan and Gijbels (2000) when using the \textit{conditional-mean} approach. We start out with the asymptotic optimal bandwidth given by formula (3.21) in Fan and Gijbels (1996, p.~68). To facilitate its practical implementation, we use the direct plug-in method proposed by Ruppert et al. (1995, pp.~1262--1263). In the latter reference, the resulting bandwidth is denoted by $\widehat{h}_{DPI}$, which in the present paper we denote by $\widehat{b}$ to avoid possible notational confusion with the estimate $\widehat{h}$ of $h$.

When using the \textit{conditional-median} approach, we follow
Yu and Jones (1997), who show that the optimal bandwidth in this case is equal to the estimate $\widehat{b}$ from the conditional-mean approach  multiplied by
\[
\left\lbrace \frac{\tau(1-\tau)}{\phi(\Phi^{-1}(\tau))^2}
\right\rbrace^{1/5} ,
\]
where $\tau =1/2$ due to our median based approach. The $\phi$ in the above quantity is the standard normal density, and $\Phi^{-1}$ is the standard normal quantile function. Hence, in summary, the optimal bandwidth under the conditional-median approach is $\widehat{b} (\pi/2)^{1/5}$.

\section{Measuring the lack of co-monotonicity}
\label{sec.diss}

In view of the above discussion, we can now assume that for any given scatterplot we have constructed a well-fitting function $\widehat{h}: [0,1] \to [0,1]$: if it happens to be increasing, then we say that the random variables $X$ and $Y$ have co-monotonic movements, but if not, then we want to assess how much the function deviates from the increasing pattern. This we accomplish using an index that takes value $0$ when $\widehat{h}$ is increasing and some positive value otherwise: the more the function deviates from the increasing pattern, the larger the value. The index, which we call the \textit{lack of co-monotonicity} (LOC) index, is discussed next.

\subsection{The LOC index}
\label{sec.index}

We start with the well-known notion of increasing rearrangement (cf., e.g., Hardy et al., 1952) which for our function $\widehat{h}: [0,1] \to [0,1]$ is defined by
\[
I_{\widehat{h}}(t)=\inf \{x\in \mathbf{R}: G_{\widehat{h}}(x) \ge t\}
\]
for all $t \in [0,1]$, where
$G_{\widehat{h}}(x)= \lambda \{ s \in [0,1]: \widehat{h}(s) \leq x \}$ and $\lambda$ is the Lebesque measure.

\begin{note}\rm
If we interpret the function $\widehat{h}$ as a random variable on the probability space $\{[0,1],\mathcal{B},\lambda\}$, then statisticians would immediately recognize that $G_{\widehat{h}}(x)$ is the cumulative distribution function of $\widehat{h}$, and $I_{\widehat{h}}(t)$ is the quantile function of $\widehat{h}$. We find these interpretations useful to work out good intuition on the subject.
\end{note}

Hence, to construct an index that would measure the lack of, or departure from, co-monotonicity between pairs of variables, we need to choose an appropriate functional that would couple $I_{\widehat{h}}$ and $\widehat{h}$ in such a way that the resulting quantity would be zero if and only if the functions $I_{\widehat{h}}$ and $\widehat{h}$ coincide, that is, the fitted function $\widehat{h}$ is increasing (to be more precise, non-decreasing). Among such candidates are the maximal distance between $I_{\widehat{h}}$ and $\widehat{h}$, called the $\sup$-norm, as well as the integrated distance between the two functions, called the $L_1$-norm. Though mathematically appealing, the two choices are not good candidates for the purpose due to the lack of so-called co-monotonic addition (to be explained in a moment) as has been pointed out by Qoyyimi and Zitikis (2014) in the $L_1$ case.

Qoyyimi and Zitikis (2014) argue that a suitable candidate for LOC index is
\begin{equation}
\mathcal{L}(\widehat{h}) = \int_0^1 t \big(I_{\widehat{h}}(t)-\widehat{h}(t)\big)dt.
\label{def.index}
\end{equation}
The integral is always non-negative, equal to $0$ for every increasing function, and takes on  (strictly) positive values for all other functions. Furthermore, $\mathcal{L}(\widehat{h}+d)=\mathcal{L}(\widehat{h})$ for every real constant $d$,
and $\mathcal{L}(c\widehat{h})=c \mathcal{L}(\widehat{h})$ for every non-negative constant $c$. If $\widehat{g}$ is a function co-monotonic with $\widehat{h}$, which means that both $\widehat{g}$  and $\widehat{h}$ increase and decrease on the same intervals of their joint domain of definition, then $\mathcal{L}(\widehat{g}+\widehat{h}) = \mathcal{L}(\widehat{g}) + \mathcal{L}(\widehat{h})$. We view this co-monotonicity property important for every LOC index to satisfy, and this is the reason we have abandoned the use of the aforementioned $\sup$- and $L_1$-norms.

Given the prominent role that the notion of monotone rearrangement is playing in the definition of the LOC index $\mathcal{L}$, it is instructive to mention that the notion has been very successfully used in quite a number of applications:
\begin{itemize}
\item
Efficient insurance contracts (e.g., Carlier and Dana, 2005;  Dana and Scarsini, 2007).
\item
Rank-dependent utility theory  (Quiggin, 1982, 1993; also Carlier and Dana, 2003, 2008, 2011).
\item
Continuous-time portfolio selection (e.g., He and Zhou, 2011; Jin and Zhou, 2008).
\item
Statistical applications such as performance improvement of estimators (e.g., Chernozhukov et al., 2009, 2010) and optimization problems (e.g., R\"{u}schendorf, 1983).
\end{itemize}
These are of course just a few illustrative topics and references, but they lead us into the vast literature on monotone rearrangements and their manifold uses.

\subsection{Computational formula}
\label{sec.empindex}

Given its properties, the LOC index $\mathcal{L}$ is attractive  but highly unwieldy even when  $\widehat{h}$ has a simple mathematical expression, let alone when it arises nonparametrically from a scatterplot. Hence, we need a simple computational method for the index even when $\widehat{h}$ does not have an explicit mathematical expression.

To this end, we first partition the interval $[0,1]$ into $m$ subintervals using the points $i/m$, $i=1,2,\dots m$. Then for each $i=1,2,\dots m$ we choose any point $t_i \in ((i-1)/m,i/m]$ for which the value $\tau_i:=\widehat{h}(t_i)$ is available. Hence, we have the step-wise function $\widehat{D}_m:[0,1]\to [0,1]$ defined by
\begin{equation}
\label{estf}
\widehat{D}_m(t)=
\begin{cases}
\tau_1 &\text{when}\quad t = 0, \\
\tau_i & \text{when}\quad  t \in \left(\frac{i-1}{m}, \frac{i}{m}\right],
\end{cases}
\end{equation}
whose LOC index has a very simple computational formula (proof in Appendix \ref{appendix})
\begin{equation}
\label{eq.estlindex}
\mathcal{L}(\widehat{D}_m) = \left(\frac{1}{m}\right)^2 \sum_{i=1}^{m} i \left(\tau_{i:m}-\tau_i\right),
\end{equation}
where $\tau_{1:m}\le \cdots \le \tau_{m:m}$ are the ordered values of $\tau_1, \dots, \tau_m$. Furthermore (proof in Appendix \ref{appendix}), when $m\to \infty $, then
\begin{equation}
\label{eq.consistent-1}
\mathcal{L}(\widehat{D}_m) \to \mathcal{L}(\widehat{h}).
\end{equation}
Hence, to calculate $\mathcal{L}(\widehat{h})$ numerically, we need to calculate $\mathcal{L}(\widehat{D}_m)$, which approximates $\mathcal{L}(\widehat{h})$ as precisely as desired provided that $m$ is sufficiently large.

\section{Data analysis and findings}
\label{sec.subjects}

\begin{figure}[h!]
	\centering
	 \subfigure[Mathematics--Reading]{\includegraphics[scale=0.25]{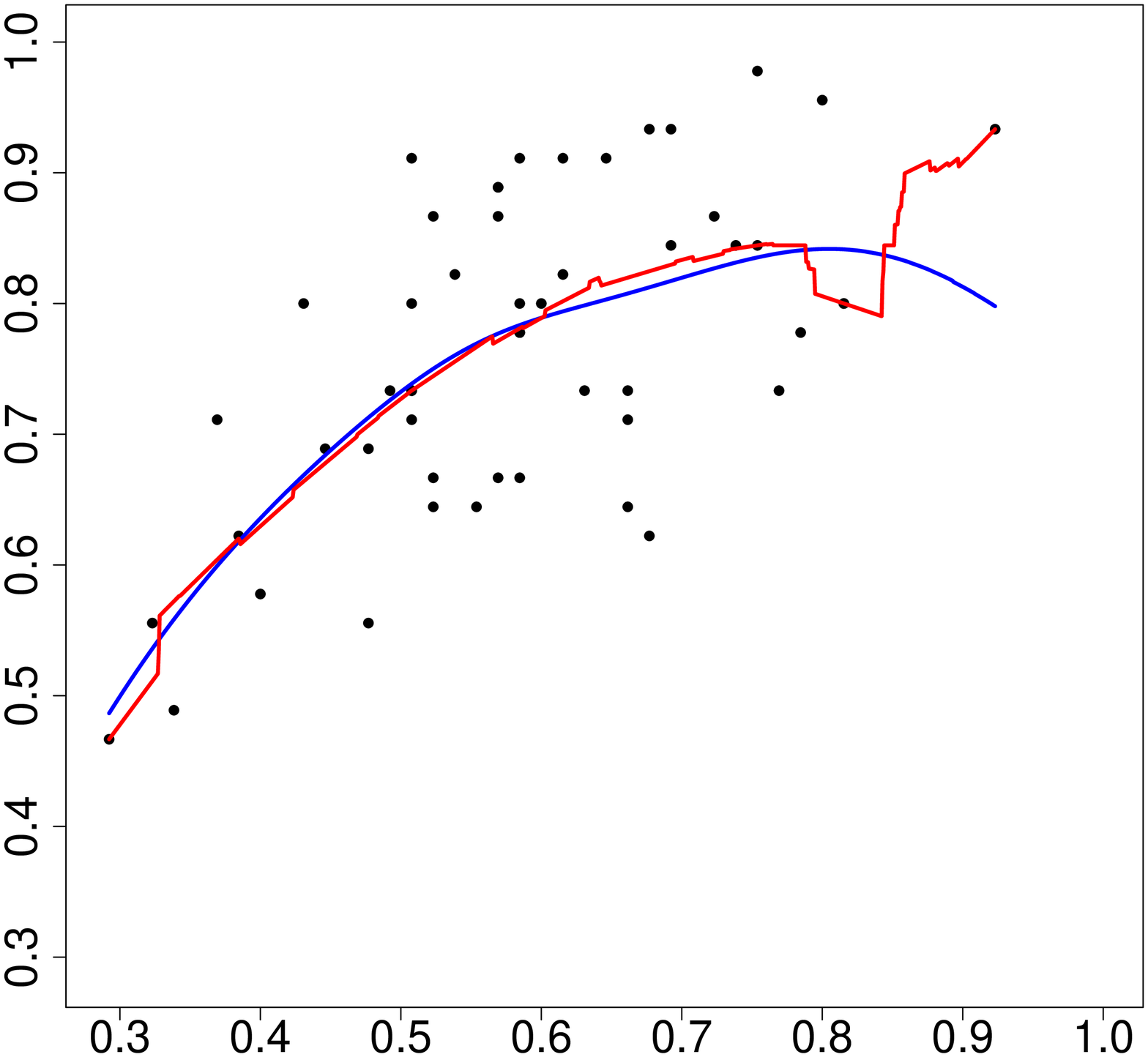}}
	\qquad \subfigure[Mathematics--Spelling]{\includegraphics[scale=0.25]{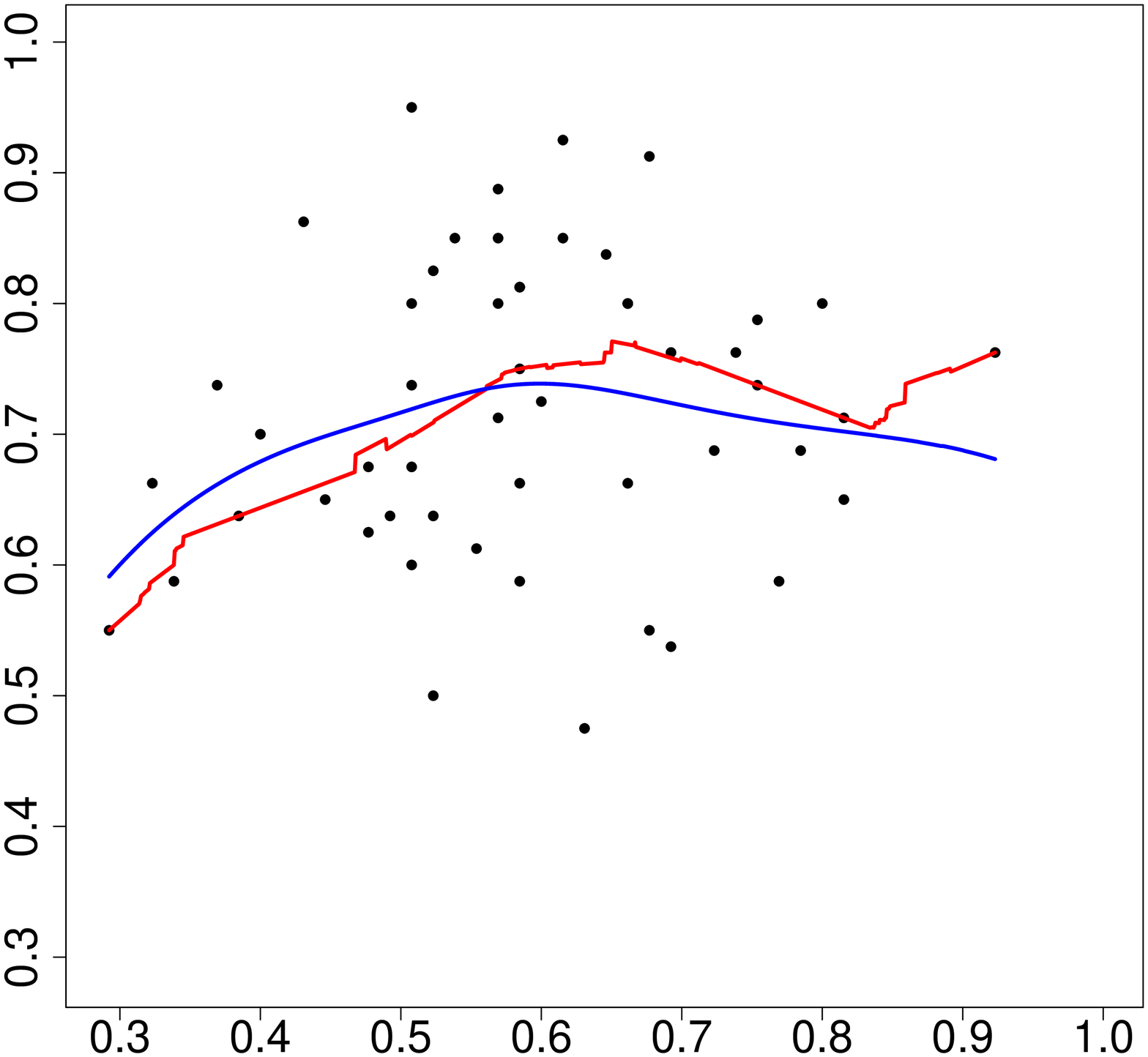}} \\
	 \subfigure[Reading--Mathematics]{\includegraphics[scale=0.25]{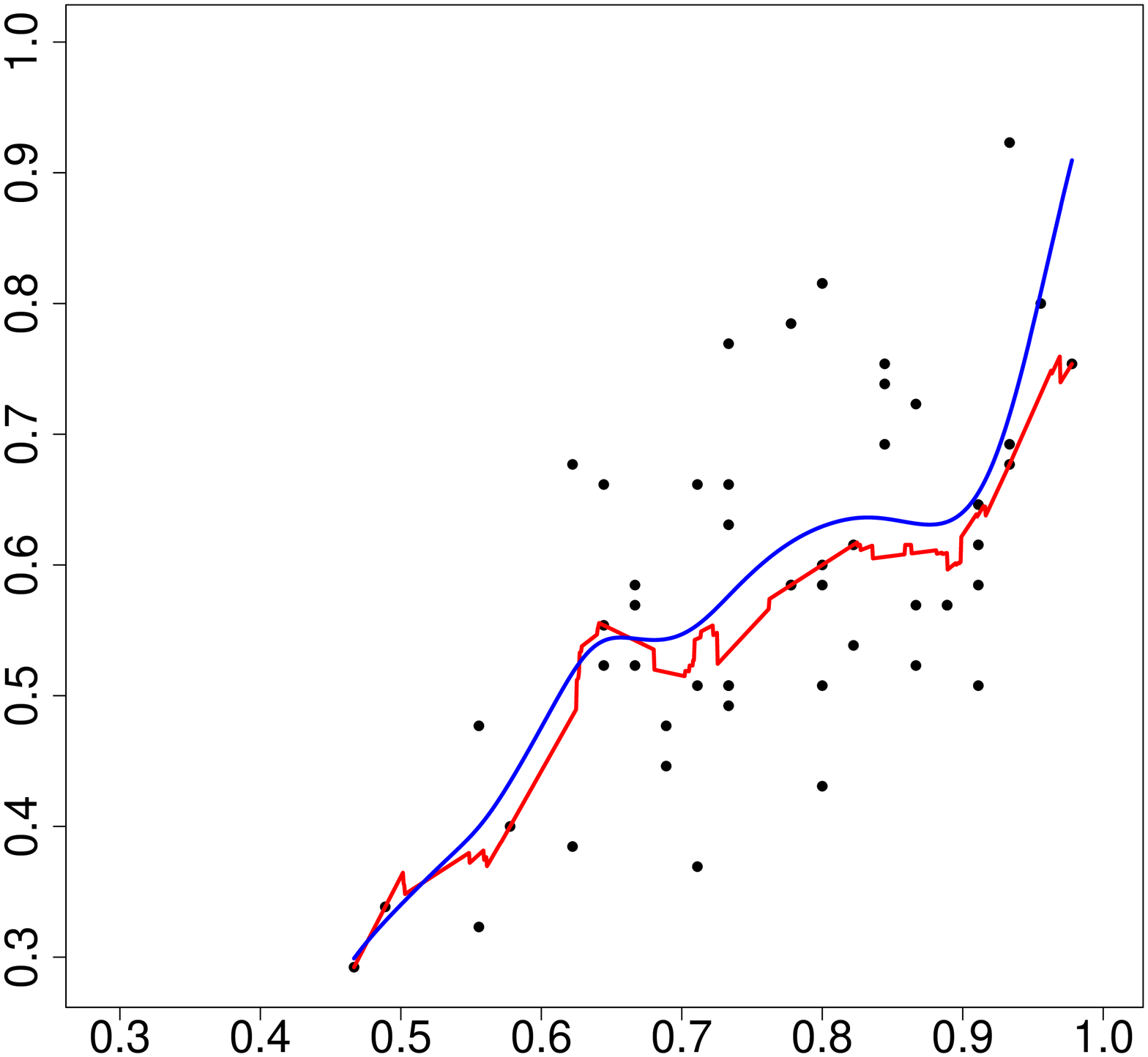}}
\qquad	 \subfigure[Reading--Spelling]{\includegraphics[scale=0.25]{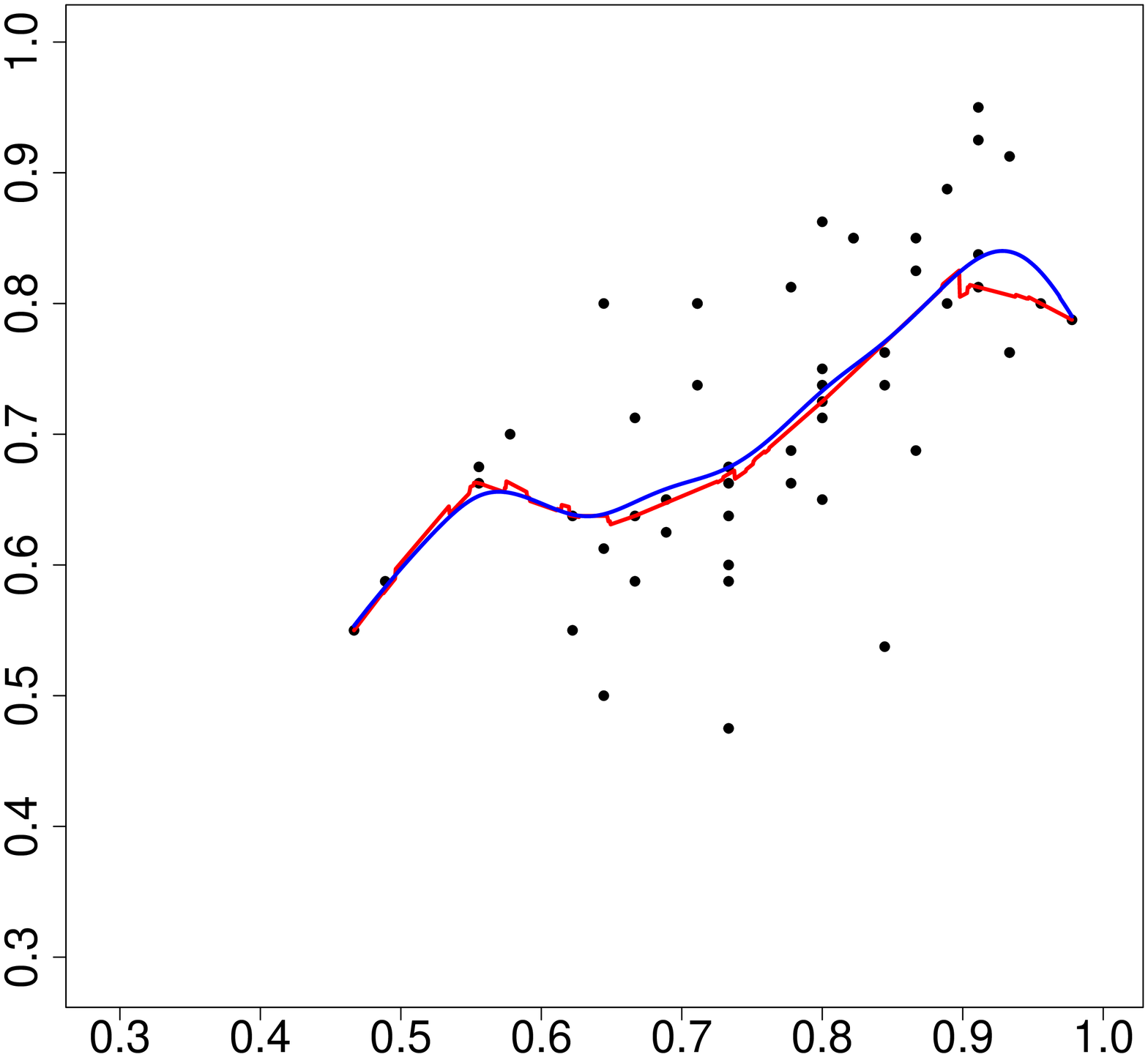}} \\
	 \subfigure[Spelling--Mathematics]{\includegraphics[scale=0.25]{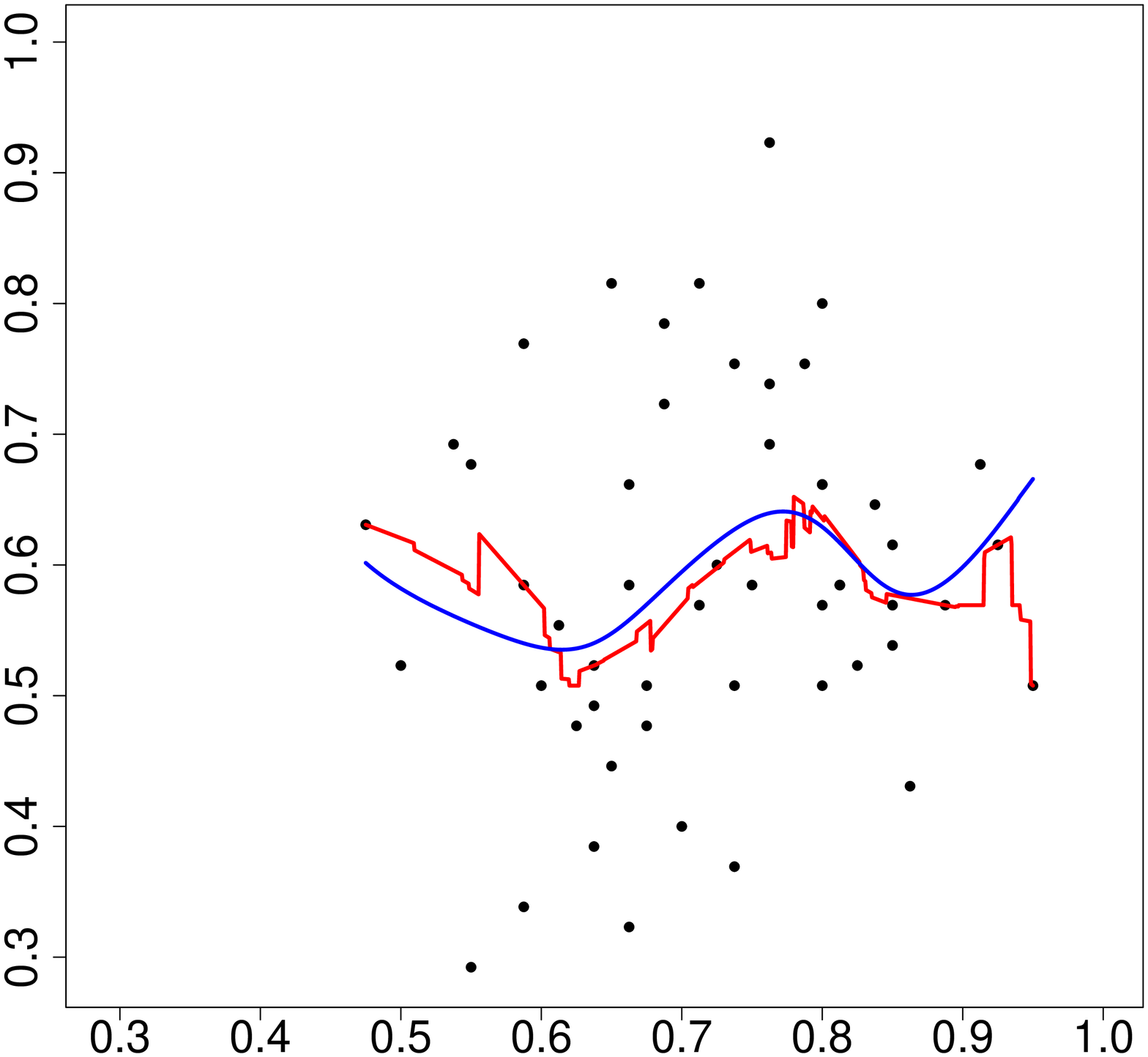}}
\qquad	 \subfigure[Spelling--Reading]{\includegraphics[scale=0.25]{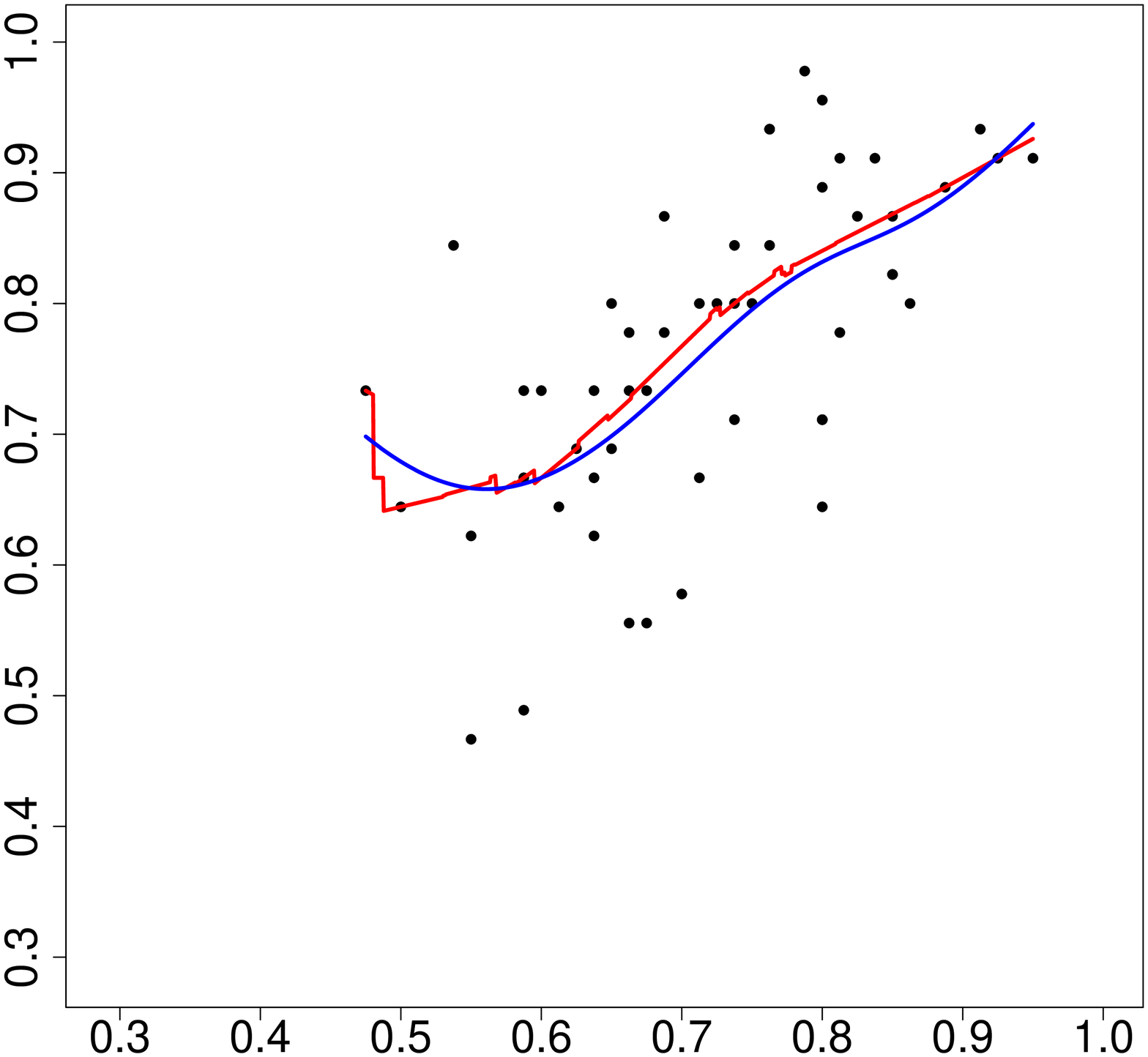}} \\
	\caption{Conditional-mean (blue) and conditional-median (red) based curves}
	\label{fig.kernandquantall}
\end{figure}

We work with six scatterplots, and to each of them we fit two curves: one using the conditional-mean approach and the other one using the conditional-median approach. In both cases, we use the same mathematical notation $\widehat{h}$ but when plotting in Figure \ref{fig.kernandquantall},
we use different colors to distinguish the two cases. The technicalities of curve fitting follow next, for which we use the R software (R Core Team, 2013).

In the case of the \textit{conditional-mean} approach, we use the local linear kernel regression method as discussed in Section \ref{sec.hhat}. To aid us with computations,  we use the R package \texttt{Kernsmooth} (Wand and Ripley, 2014) with the function \texttt{dpill} assigned for selecting the optimal bandwidth and the function \texttt{locpoly} (with degree=1) for curve fitting. We set the grid size to 1,000.

In the case of the \textit{conditional-median} approach, we use the R package \texttt{quantreg} (Koenker, 2015) with the function \texttt{lpqr} used to obtain $\widehat{h}$ with $\tau=1/2$ and $m=1,000$. We see from the six panels of Figure \ref{fig.kernandquantall} that all the estimates $\widehat{h}$ are more jiggly than those arising from the conditional-mean approach. Definitely, we can improve them with more work and a more sophisticated tuning of the parameters, but this would beat our purpose of showing that we can easily calculate the LOC index irrespective of how much irregular the function is.

Based on our visual assessment, no function in Figure \ref{fig.kernandquantall} appears to be increasing over its entire domain of definition. Nevertheless, we may argue that some of them are more increasing than others. To substantiate this claim, we employ the LOC index discussed in Section \ref{sec.diss}. The following terminology is useful.

\begin{definition}\rm
Given two functions $\widehat{g},\widehat{h}:[0,1]\to [0,1]$, we say that
\begin{enumerate}[(1)]
  \item
  $\widehat{g}$ deviates from increasing pattern by the amount $\mathcal{L}(\widehat{g})$;
  \item
  $\widehat{g}$ deviates less from increasing pattern than  $\widehat{h}$ when $\mathcal{L}(\widehat{g}) < \mathcal{L}(\widehat{h})$; and
  \item \label{def-3}
  pairs $(v_i,w_i)$, $i=1,\dots, n$,  exhibit less LOC than pairs $(x_i,y_i)$, $i=1,\dots, m$,  when $\mathcal{L}(\widehat{g}) < \mathcal{L}(\widehat{h})$, where $\widehat{g}$ arises from the pairs $(v_i,w_i)$  and $\widehat{h}$ from $(x_i,y_i)$.
\end{enumerate}
\label{def.ldeviate}
\end{definition}

Following the guidelines of Section \ref{sec.empindex}, we produce the step-wise approximation $D_m$ of the function $\widehat{h}$. Then we calculate the index $\widehat{\mathcal{L}}(\widehat{D}_m) $ according to formula (\ref{eq.estlindex}). Findings in the form of 'LOC matrices' are presented in Tables \ref{tab.dismatmean} and \ref{tab.dismatmed},
\begin{table}[h!]
\centering
\begin{tabular}{l c c c}
	\hline
	              & \multicolumn{3}{c}{$Y$} \\ \cline{2-4}
	\quad $X$ & Mathematics  & Reading & Spelling \\
	\hline Mathematics & 0.000000 & 0.231814 & 1.759735 \\
	Reading & 0.007202 & 0.000000 & 0.097565 \\
	Spelling & 0.855971 & 0.145532 & 0.000000 \\
	\hline
\end{tabular}
\caption{Conditional-mean based LOC matrix (entries multiplied by $1,000$).}
\label{tab.dismatmean}
\end{table}
\begin{table}[h!]
	\centering
	\begin{tabular}{lccc}
		\hline
						& \multicolumn{3}{c}{$Y$} \\ \cline{2-4}
		\quad $X$  & Mathematics  & Reading & Spelling \\
		\hline Mathematics & 0.000000 & 0.286703 & 0.923108 \\
		Reading & 0.007911 & 0.000000 & 0.163541 \\
		Spelling & 2.197968 & 0.175055 & 0.000000 \\
		\hline
	\end{tabular}
	\caption{Conditional-median based LOC matrix (entries multiplied by $1,000$).}
	\label{tab.dismatmed}
\end{table}
whose entries are the values of the LOC index: the larger the value, the more the corresponding pairs deviate from the co-monotonic pattern.

The LOC matrix is, naturally, asymmetric, and it should be such in order to match the asymmetry that we see in the respective paired panels of Figure \ref{los-fitting}. For example, the entry $0.231814$ in Table \ref{tab.dismatmean} is the value (multiplied by $1,000$) of the LOC index for Mathematics-Reading, whereas $0.007202$ is the value (multiplied by $1,000$) of the LOC index for Reading-Mathematics. We have multiplied all the original LOC-index values by $1,000$ to avoid recording too many decimal zeros in the tables.

Naturally, one may also wish to know how much a given study subject influences the other ones, which leads us in the direction of causality (cf., e.g., Pearl, 2009; and references therein), which at this stage of our research we want to avoid discussing. Nevertheless, the reader may wish to draw some conclusions from Tables \ref{tab.dismatmean} and \ref{tab.dismatmed}, as well as from the scatterplots of Figure \ref{fig.kernandquantall}. Note that even though the corresponding entries of Tables \ref{tab.dismatmean} and \ref{tab.dismatmed} are different, the causality-type conclusions that we may infer from both of them would not contradict each other. This may not always be the case, especially if data are considerably skewed. In the case of the data that we are exploring, however, the descriptive statistics and histograms in Section \ref{sec.data} suggest fairly symmetric distributions of all the three study subjects.

\section{Comparing the LOC index with Liebscher's $\zeta$ }
\label{sec.liebscher}

Here we compare the LOC index $\mathcal{L}$ with Liebscher's (2014) coefficient $\zeta_{X,Y}$ of monotonically increasing dependence.  Naturally, to understand $\zeta_{X,Y}$ we only need to understand its expectation-based part, which under the quadratic function $\psi(x)=x^2/2$ is equal to
\[
I_{X,Y}={1\over 2}\mathbb{E}\Big [\big (F(X)-G(Y)\big )^2 \Big ] . 
\]

\begin{note}\label{note-2}\rm
The quantity $I_{X,Y}$ is closely related to the Spearman rank correlation coefficient, denoted here by $S_{X,Y}$, which is, by definition, equal to the Pearson correlation coefficient between $F(X)$ and $G(Y)$. Hence, we easily check the equation $S_{X,Y}= 1-12\,I_{X,Y}$.
\end{note}

Next we work with a scatterplot $(x_{i}, y_{i})$, $i=1,\dots , n$, which we view as our `population.' To avoid computational complications that inevitably arise when dealing with ranks when some of the $x_{i}$'s or $y_{i}$'s are equal, throughout the rest of this section we work under the assumption
\begin{equation}\label{assumption-0}
\textrm{$x_{i}\ne x_j$ and $y_{i}\ne y_j$ whenever $i\ne j$.}
\end{equation}

\begin{note}\label{note-3}\rm
Assumption (\ref{assumption-0}) is violated by the data-set of Thorndike and Thorndike-Christ (2010). However, this is not an issue because we can always add negligible noise (e.g., independent and identically distributed normal random variables with means $0$ and very small standard deviations, say $10^{-5}$) and make all the marks unequal without \textit{practically} changing their numerical values.
\end{note}

Let $F_n$ and $G_n$ be the marginal cdf's defined by $F_n(x)=\sum_{i=1}^n \mathbf{1}\{x_i\le x\}/n$ and $G_n(x)=\sum_{i=1}^n \mathbf{1}\{y_i\le x\}/n$. Under this `finite population' scenario, the quantity $I_{X,Y}$ becomes
\[
I_{n,x,y}=\frac{1}{2n} \sum_{i=1}^n(F_n(x_i)-G_n(y_i))^2 .
\]
Let $x_{1:n}< \cdots < x_{n:n}$ be the ordered values of $x_1,\dots, x_n$, and let $y_{(1)}, \dots , y_{(n)}$ be the corresponding induced ordered values. In other words, the original pairs $(x_i, y_i)$ have been ordered according to their first coordinates and the resulting pairs are now $(x_{i:n}, y_{(i)})$. With the notation
\begin{equation}\label{eq-99}
r_i=nG_n(y_{(i)})
\end{equation}
we have
\begin{align}
I_{n,x,y}
&=\frac{1}{2n} \sum_{i=1}^n(F_n(x_{i:n})-G_n(y_{(i)}))^2
\notag
\\
&= \frac{1}{2n}\sum_{i=1}^n \left(\frac{i}{n}-\frac{r_i}{n}\right)^2
\notag
\\
&= \frac{1}{2n^3}\sum_{i=1}^n \left( i- r_i\right)^2,
\label{def.spearman}
\end{align}
where we used the equation $F_n(x_{i:n})=i/n$.

Next we construct a function $h_n^0:[0,1]\to [0,1]$ such that $\mathcal{L}(h_n^0)$ is equal to the right-hand side of equation (\ref{def.spearman}) or, in other words, such that
\begin{equation}\label{eq-100}
\mathcal{L}(h_n^0)=I_{n,x,y} .
\end{equation}
Namely, for every $i=1,\dots, n$, let
\begin{equation}\label{eq-101}
h_n^0(t)={r_i\over n} \quad \textrm{for all} \quad t \in \bigg ( {i-1\over n},{i\over n} \bigg ].
\end{equation}
The LOC index of the function $h_n^0$ is
\begin{align}
\mathcal{L}(h_n^0)&=\sum_{i=1}^n \left(\frac{i}{n}-\frac{r_i}{n}\right)\int_{(i-1)/n}^{i/n}tdt
\notag
\\
&=\frac{1}{2n^3}\sum_{i=1}^n (i-r_i)(2i-1)
\notag
\\
&=\frac{1}{2n^3}\sum_{i=1}^n (i-r_i)^2,
\label{lrank}
\end{align}
where we used the equations $\sum_{i=1}^n i=\sum_{i=1}^n r_i$ and $\sum_{i=1}^n i^2=\sum_{i=1}^n r_i^2$. This establishes equation (\ref{eq-100}) and helps us to connect the LOC index $\mathcal{L}$ with Liebscher's $\zeta$.

\begin{figure}[h!]
	\centering
	\subfigure[Raw data]{\includegraphics[scale=0.25]{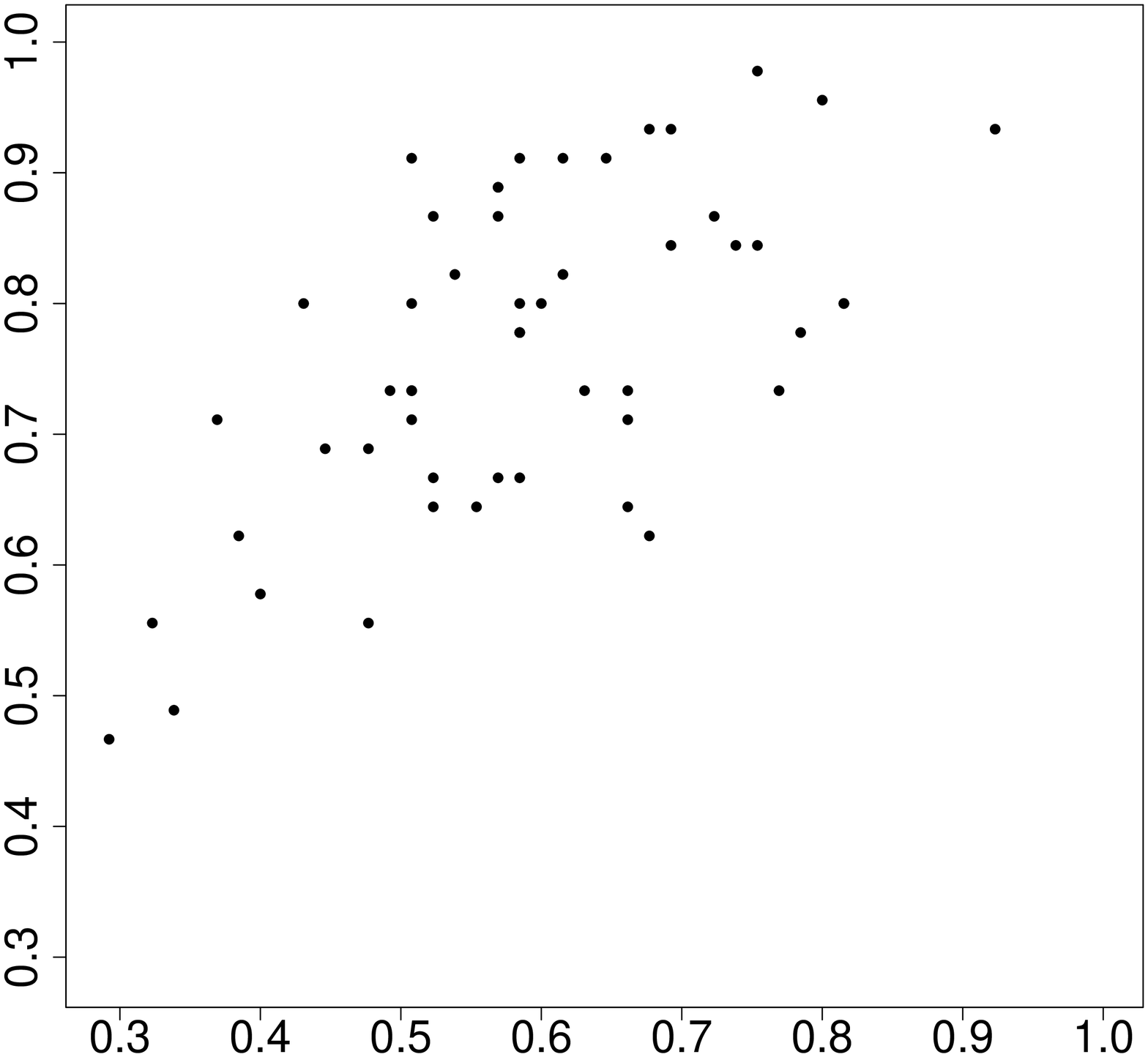}}\qquad
	\subfigure[Ranks]{\includegraphics[scale=0.25]{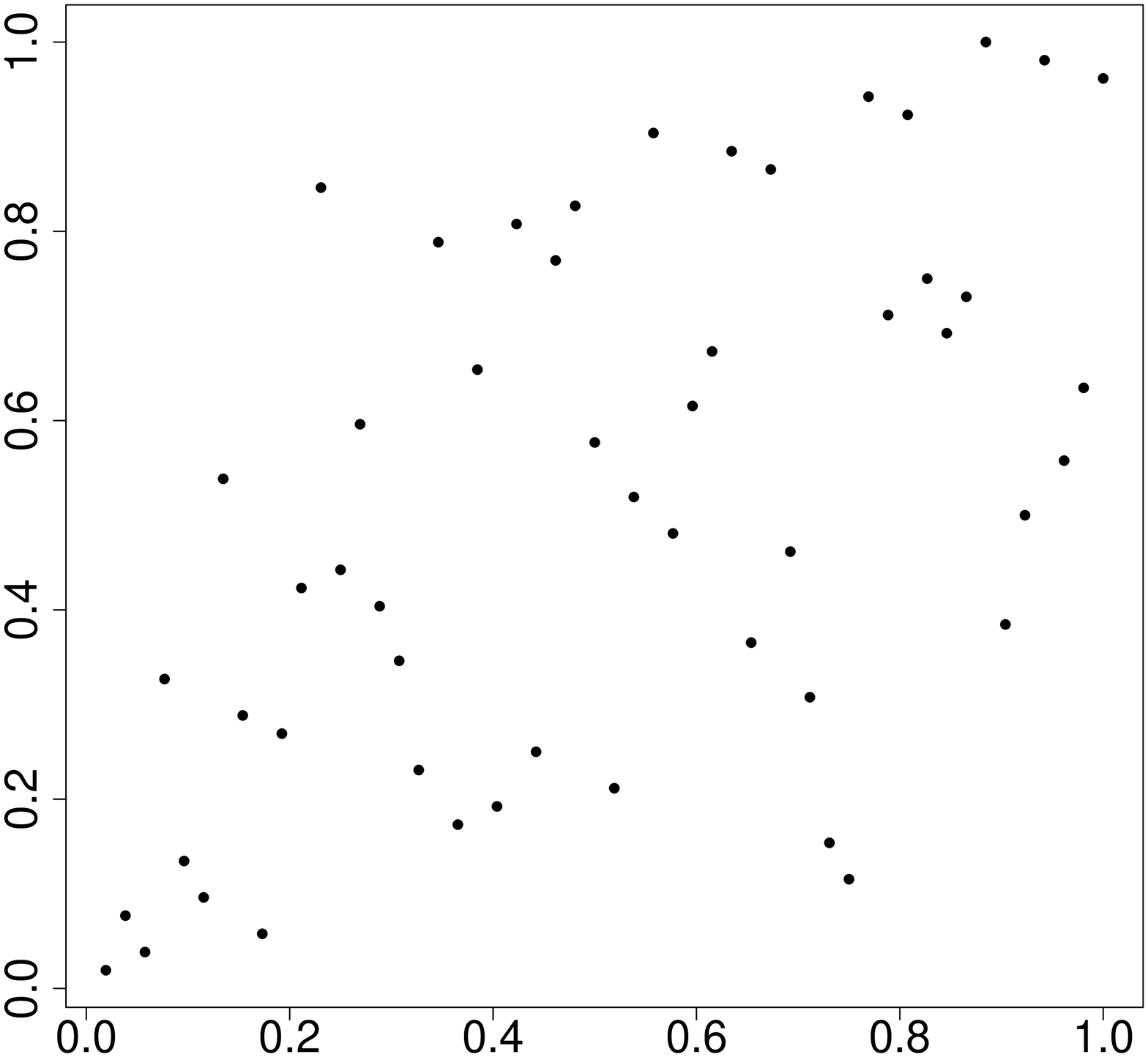}}\\
	\subfigure[$h_n$ (blue) and $I_{h_n}$ (red)]{\includegraphics[scale=0.25]{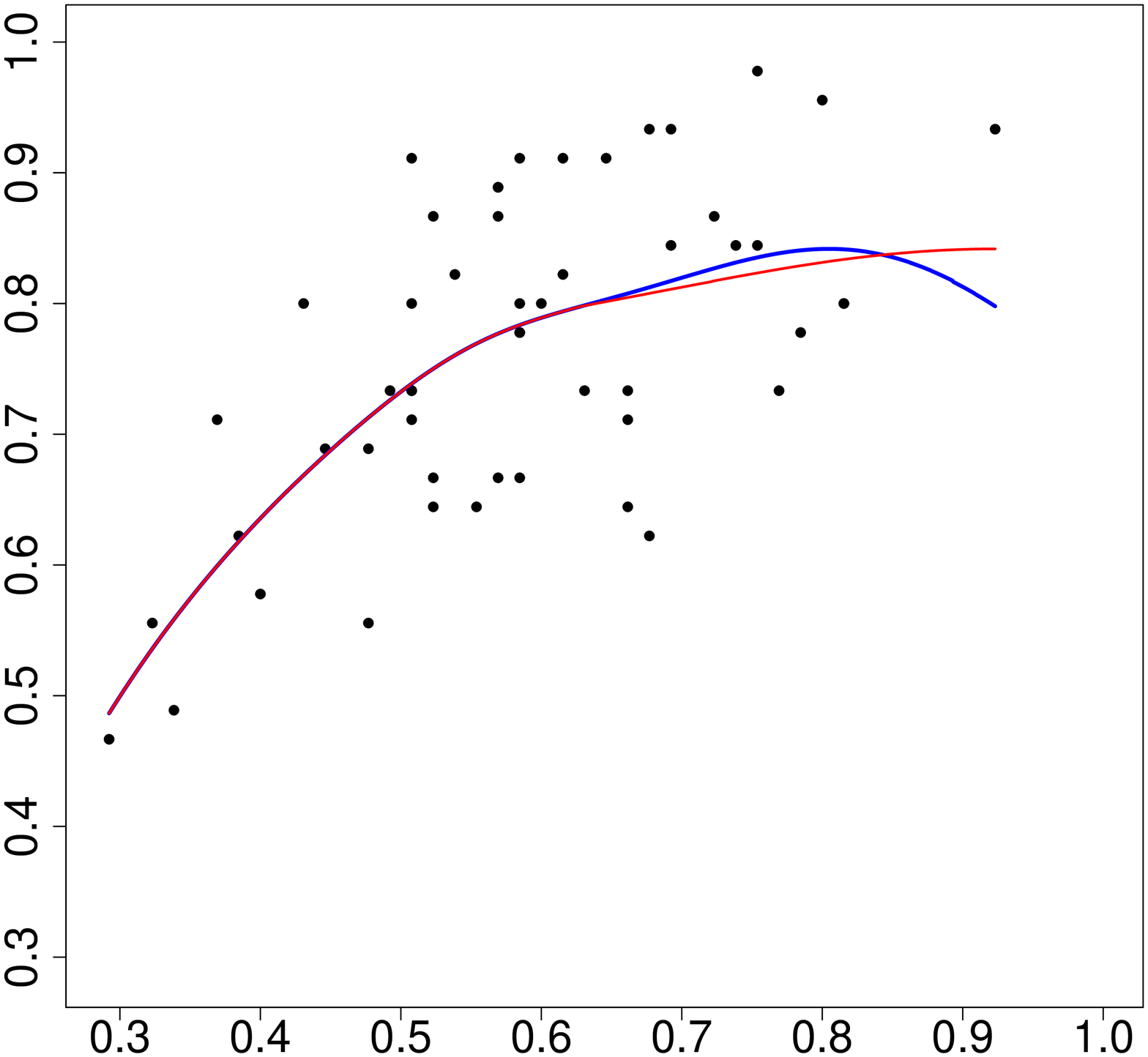}}\qquad
	\subfigure[$h_n^0$ (blue) and $I_{h_n^0}$ (red)]{\includegraphics[scale=0.25]{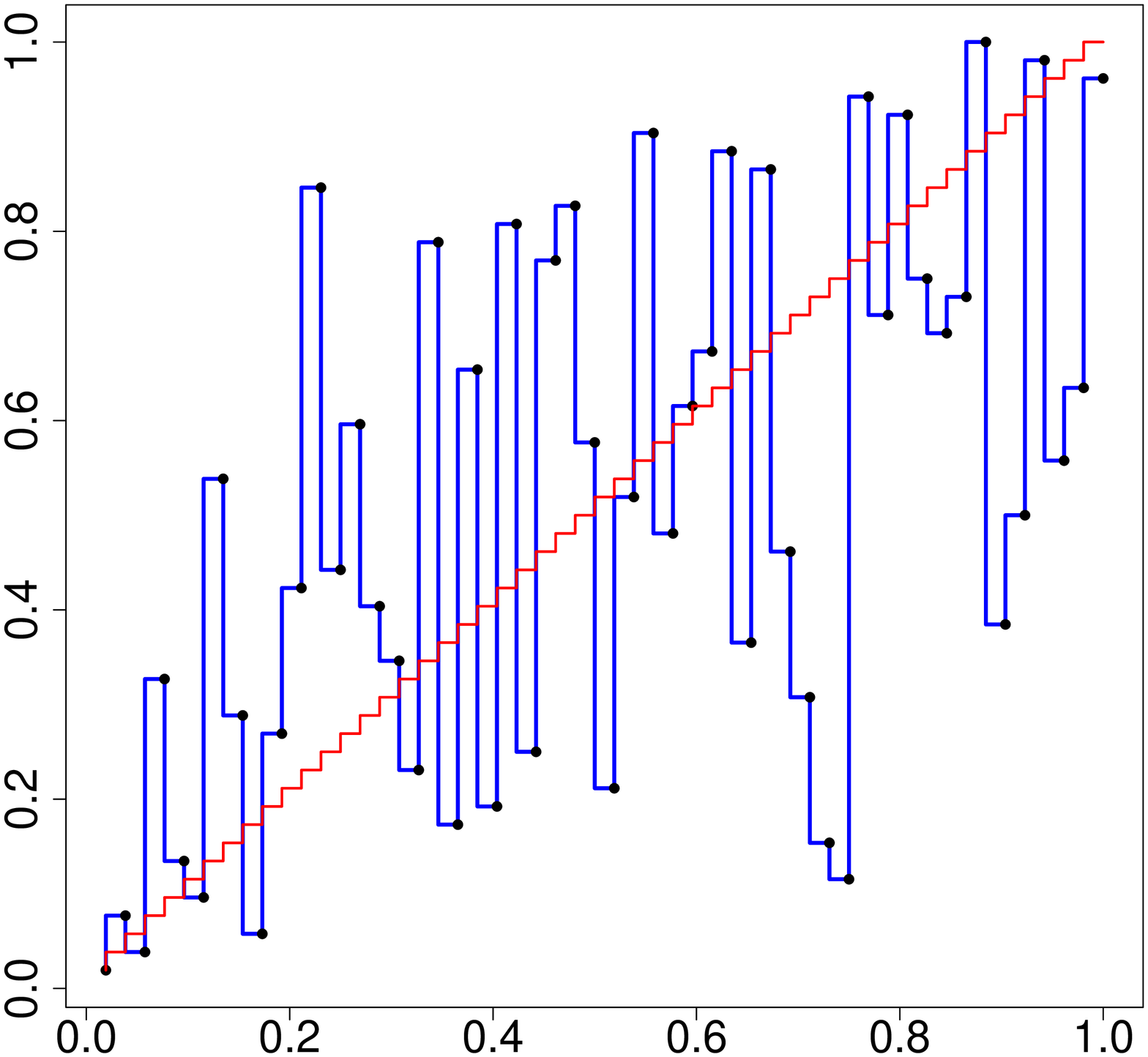}}
	\caption{Raw and rank-based scatterplots (with added negligible noise) and fitted functions}
	\label{fig.rankdiss}
\end{figure}

For this, we first observe that the set of equations $h_n^0(i/n)=r_i/n$, $i=1,\dots , n$, is equivalent to the set $h_n^0(F_n(x_{i:n}))=G_n(y_{(i)})$, $i=1,\dots , n$, which is in turn equivalent to the set of equations  $h_n^0(F_n(x_{i}))=G_n(y_i)$), $i=1,\dots , n$. This implies that  Liebscher's $\zeta$ is the LOC index $\mathcal{L}$ of the step-wise function $h_n^0$, which originates from the \textit{rank-based} scatterplot $(F_n(x_i),G_n(y_i))$ and not from the original scatterplot $(x_{i:n}, y_{(i)})$. This also explains a considerable difference between the meanings of the two indices. To support our conclusions, we have depicted the two scenarios in Figure \ref{fig.rankdiss},
where we have used Mathematics (with added small noise; recall Note \ref{note-3}) as the `explanatory'  variable and Reading (with added small noise) as the `response.'

Consequently, in order to decide whether the problem at hand would be better served by the LOC index $\mathcal{L}$ or Liebscher's $\zeta$, we first need to decide whether the solution of the problem should rely on the original scatterplot $(x_{i:n}, y_{(i)})$, $i=1,\dots , n$, or on the rank-based scatterplot  $(F_n(x_{i:n}), G_n(y_{(i)}))$; the latter is of course equivalent to the scatterplot $(i/n, r_i/n)$, $i=1,\dots , n$. If the association between student \textit{rankings} according to their marks is of primary interest, with no consideration to causality, then Liebscher's $\zeta$ is an appropriate index. If, however, the \textit{marks} themselves are of primary interest, as is the case in the current paper, and keeping in mind that the marks are not interchangeable random variables with respect to causality, then we should rely on the original scatterplot  $(x_{i:n}, y_{(i)})$, $i=1,\dots , n$, and use the LOC index $\mathcal{L}$.

Note, however, that at present the LOC index is available only for pairs of variables, which is of immediate interest for educational psychologists, whereas Liebscher's methodology has been extended to the multivariate case (Section 5 in Liebscher, 2014) and could provide further valuable insights into the problem when all study subjects are viewed as integral parts of one `study portfolio.'

\section{Concluding notes and further work}
\label{sec.conclusion}

The herein proposed index for measuring the lack of co-monotonicity (LOC) between pairs of variables is capable of measuring the extent to which the variables deviate from co-monotonic patterns. The LOC index is designed to work with all relationships, including non-linear and non-monotonic. The performance of the index has been illustrated using the Thorndike and Thorndike-Christ (2010) data-set consisting of student marks on three study subjects.

In addition to the educational assessment problem that we have tackled in this paper, there are of course numerous other applications where monotonicity, or lack of it, matters, and we next present a few examples to illustrate the point.

The presence of a deductible $d\ge 0$ often changes the profile of insurance losses (e.g., Brazauskas et al., 2015). Because of this and other reasons, given two losses $X$ and $Y$, which may not be observable, decision makers may wish to determine whether the observable losses $X_d=[X\mid X>d]$ and $Y_d=[Y\mid Y>d]$ are stochastically (ST) ordered, say $X_d \le_{\textrm{ST}} Y_d$ for every $d\ge 0$. It is well known that this ordering, which is known in the literature as the hazard rate ordering,  is equivalent to determining whether the ratio $S_Y(x)/S_X(x)$ of $X$ and $Y$ survival functions is non-decreasing in $x$.

More generally, one may wish to determine whether for every deductible $d\ge 0$ and every policy limit $L>d$, the observable insurance losses $X_{d,L}=[X\mid d\le X \le L ]$ and $Y_{d,L}=[Y\mid d\le Y \le L ]$ are stochastically ordered. This ordering, which is known in the literature as the likelihood ratio ordering, is equivalent to determining whether the ratio $f_Y(x)/f_X(x)$  of $X$ and $Y$ density functions is non-decreasing in $x$. For further details on various stochastic orderings and their manifold uses, we refer to Shaked and Shanthikumar (2006), Li and Li (2013), and references therein.

We next briefly present a few more examples and related references where monotonicity, or lack of it, of certain functions plays an important role:
\begin{itemize}
  \item
  Growth curves (cf., e.g., Chernozhukov et al., 2009; Panik, 2014).
  \item
  Mortality curves  (cf., e.g., Gavrilov and Gavrilova, 1991; Bebbington et al., 2011).
  \item
  Positive regression dependence and risk sharing (cf., e.g., Lehmann,  1966; Dana and Scarsini, 2007).
  \item
  Comonotonicity, portfolio construction, and capital allocations (cf., e.g., Dhaene et al., 2006; Furman and Zitikis, 2008).
  \item
  Decision theory and stochastic ordering (cf., e.g., Denuit et al., 2005; Shaked and Shanthikumar, 2006; Li and Li, 2013).
  \item
  Engineering reliability and risks (cf., e.g., Lai and Xie, 2006; Li and Li, 2013).
\end{itemize}

One unifying feature of these diverse works is that they impose monotonicity requirements on certain functions, which are generally unknown, and thus researchers may seek for statistical models and data for determining their shapes. To illustrate the point, we recall, for example, the work of Bebbington et al. (2011) who specifically set out to determine whether mortality continues to increase or starts to decelerate after a certain species related late-life age. This is known in the literature as the late-life mortality deceleration phenomenon. Hence, we can rephrase the phenomenon as a question: is the mortality function always increasing? Naturally, we do not elaborate on this topic any further in this paper, referring the interested reader to Bebbington et al. (2011), and references therein.

To verify the monotonicity of functions such as those noted in the above examples, researchers quite often assume that the functions belong to some parametric or semiparametric families. One may not, however, be comfortable with this element of subjectivity and thus prefers to rely solely on data to make a judgement. Under these circumstances, verifying monotonicity becomes a non-parametric problem, whose solution asks for an index that, for example, takes on the value $0$ when the function under consideration is non-decreasing and on positive values otherwise. This is exactly the topic that we have dealt with in the present paper.

\appendix
\section{Appendix: proofs}
\label{appendix}

\begin{proof}[Proof of equation (\ref{eq.estlindex})]
We check that
\[
G_{\widehat{D}_m}(x)=\mathbf{1}_{\{\tau_{m:m}\}} + \sum_{i=1}^{m} \frac{i-1}{m} \mathbf{1}_{[\tau_{i-1:m},\tau{i:m})}(t)
\]
and
\[
I_{\widehat{D}_m}(t)=\sum_{i=1}^{m} \tau_{i:m} \mathbf{1}_{((i-1)/m,i/m]}(t).
\]
Hence,
\begin{align*}
\mathcal{L}(\widehat{D}_m)
& = \int_0^1 t \left(I_{\widehat{D}_m}(t)-\widehat{D}_m(t)\right)dt
\\
&=
\sum_{i=1}^{m}(\tau_{i:m}-\tau_i)\int_0^1 t \mathbf{1}_{((i-1)/m,i/m]}(t)dt
\\
&=\sum_{i=1}^{m}(\tau_{i:m}-\tau_i) \left(\frac{i}{m^2}-\frac{1/2}{m^2}\right)
\\
&=\left(\frac{1}{m}\right)^2\sum_{i=1}^{m}i(\tau_{i:m}-\tau_i).
\end{align*}
This concludes the proof of equation (\ref{eq.estlindex}).
\end{proof}

\begin{proof}[Proof of statement (\ref{eq.consistent-1})]
For any two integrable functions $u, v: [0,1] \to \mathbb{R}$, the intergral $ \int_0^1 |I_u(t)-I_v(t)|dt $ does not exceed $ \int_0^1 |u(t)-v(t)|dt$ (e.g., Denneberg, 1994).
Setting $u=\widehat{D}_m$ and $v=\widehat{h}$, and using the triangle inequality, we obtain the bound
\[
\big \vert \mathcal{L}(\widehat{D}_m) - \mathcal{L}(\widehat{h}) \big \vert \le 2\int_0^1 |\widehat{D}_m(t)-\hat{h}(t)|dt.
\]
Since the function $\widehat{h}$ is finite and integrable, the right-hand side of the above bound converges to zero when $m\to \infty $. This finishes the proof of statement (\ref{eq.consistent-1}).
\end{proof}

\section*{Acknowledgments}

We are indebted to the referees and editors for constructive criticism and numerous suggestions that helped us during several revisions of the paper. Many people have commented on our results since their posting on the SSRN and arXiv technical-report repositories in February 2015 -- it would be impossible to mention all of them here but we thank them all most sincerely. The first author gratefully acknowledges his PhD study support by the Directorate General of Human Resources for Science, Technology and Higher Education, Indonesia. The second author has been supported by the Natural Sciences and Engineering Research Council (NSERC) of Canada.


\begin{thebibliography}{99}


%
%
%
%
%
%
%



\bibitem{blz2011}
Bebbington, M., Lai, C.D. and Zitikis, R. (2011).
Modelling deceleration in senescent mortality.
\textit{Mathematical Population Studies} \textbf{18} (1), 18--37.


%
%
%



\bibitem{bjz2015}
Brazauskas, V., Jones, B.L. and Zitikis, R. (2015).
Trends in disguise.
\textit{Annals of Actuarial Science} \textbf{9} (1), 58--71.



\bibitem{cd2003a}
Carlier, G. and Dana, R.A. (2003).
Pareto efficient insurance contracts when the insurer's cost function is discontinuous.
\textit{Economic Theory} \textbf{21} (24), 871--893.


\bibitem{cd2005}
Carlier, G. and Dana, R. (2005).
Rearrangement inequalities in non-convex insurance models.
\textit{Journal of Mathematical Economics} \textbf{41} (4-5), 485--503.


\bibitem{cd2008}
Carlier, G. and Dana, R. (2008).
Two-persons efficient risk-sharing and equilibria for concave law-invariant utilities.
\textit{Economic Theory} \textbf{36} (2), 189--223.


\bibitem{cd2011}
Carlier, G. and Dana, R. (2011).
Optimal demand for contingent claims when agents have law invariant utilities.
\textit{Mathematical Finance} \textbf{21} (2), 169--201.






\bibitem{cfg2009}
Chernozhukov, V., Fernandez-Val, I. and Galichon, A. (2009).
Improving point and interval estimators of monotone function by rearrangement. \textit{Biometrika} \textbf{96}, 559--575.


\bibitem{cfg2010}
Chernozhukov, V., Fernandez-Val, I. and Galichon, A. (2010).
Quantile and probability curves without crossing.
\textit{Econometrica} \textbf{78} (3), 1093--1125.

%



\bibitem{ds2007}
Dana, R. and Scarsini, M. (2007).
Optimal risk sharing with background risk.
\textit{Journal of Economic Theory} \textbf{133} (1), 152--176.

\bibitem{d1994}
Denneberg, D. (1994).
{\it Non-additive Measure and Integral}. Kluwer, Dordrecht.

%
%

%


\bibitem{dlsv2012}
Dhaene, J., Linders, D., Schoutens, W. and Vyncke, D. (2012). The herd behavior index: a new measure for the implied degree of co-movement in stock markets. \textit{Insurance: Mathematics and Economics} \textbf{50} (3), 357--370.


\bibitem{dlsv2014}
Dhaene J., Linders D., Schoutens W. and Vyncke D. (2014). A multivariate dependence measure for aggregating risks. \textit{Journal of Computational and Applied Mathematics} \textbf{263} (1), 78--87.

\bibitem{dvgktv2006}
Dhaene, J., Vanduffel, S., Goovaerts, M.J., Kaas, R., Tang, Q. and Vyncke, D. (2006).
Risk measures and comonotonicity: a review.
\textit{Stochastic Models} \textbf{22} (4), 573--606.




%
%


\bibitem{f1992}
Fan, J. (1992).
Design-adaptive nonparametric regression. {\it Journal of the American Statistical Association} \textbf{87} (420), 998--1004.

\bibitem{fg1996}
Fan, J. and Gijbels, I. (1996).
{\it Local Polynomial Modelling and Its Applications}. Chapman \& Hall, London, UK.

\bibitem{fg2000}
Fan, J. and Gijbels, I. (2000).
Local polynomial fitting. In: Schimek, M.G. (ed.), {\it Smoothing and Regression: Approaches, Computation, and Application}. Wiley, New York, 229--276.




\bibitem{fz2008}
Furman, E. and Zitikis, R. (2008).
Weighted risk capital allocations.
\textit{Insurance: Mathematics and Economics} \textbf{43} (2), 263--269.




\bibitem{gg1991}
Gavrilov, L.A. and Gavrilova, N.S. (1991).
\textit{The Biology of Life Span: A Quantitative Approach}.
Harwood Academic Publishers, New York.





\bibitem{hlp1934}
Hardy, G.H., Littlewood, J.E. and P\'{o}lya, G. (1952).
{\it Inequalities} (2nd edition). Cambridge University Press, Cambridge.



\bibitem{hz2010}
He, X. and Zhou, X. (2011).
Portfolio choice via quantile.
\textit{Mathematical Finance} \textbf{21} (2), 203--231.


\bibitem{jdh2013}
Jaworski, P., Durante, F. and H\"{a}rdle, W. (2013). \textit{Copulae in Mathematical and Quantitative Finance.} Springer, Berlin.

\bibitem{jdhr2010}
Jaworski, P., Durante, F., H\"{a}rdle, W. and Rychlik, T. (2010). \textit{Copula Theory and Its Applications.} Springer, Berlin.


\bibitem{jz2008}
Jin, H. and Zhou, X. (2008).
Behavioral portfolio selection in continuous time.
\textit{Mathematical Finance} \textbf{18} (3), 385--426.

%
%



\bibitem{ks2011}
Koch, I. and  De Schepper, A. (2011). Measuring comonotonicity in m-dimensional vectors. \textit{ASTIN Bulletin} \textbf{41} (1), 191--213.



\bibitem{k2005}
Koenker, R. (2005).
{\it Quantile Regression}. Cambridge University Press, New York.

\bibitem{k2015}
Koenker, R. (2015). {\it quantreg: Quantile Regression}. R package version 5.11, URL  http://cran.r-project.org/web/packages/quantreg/index.html




\bibitem{lx2006}
Lai, C. D. and Xie, M. (2006).
\textit{Stochastic Ageing and Dependence for Reliability}. Springer, New York.



\bibitem{l1966}
Lehmann, E.L. (1966).
Some concepts of dependence.
\textit{Annals of Mathematical Statistics} \textbf{37} (5), 1137--1153.




\bibitem{ll2013}
Li, H. and Li, X. (2013).
\textit{Stochastic Orders in Reliability and Risk: In Honor of Professor Moshe Shaked}. Springer, New York.


\bibitem{l2014}
Liebscher, E. (2014).
Copula-based dependence measures. {\it Dependence Modeling}  {\bf 2} (1), 49--64.






\bibitem{n2006}
Nelsen, R.B. (2006).
{\it An Introduction to Copulas.} (Second Edition.) Springer, New York.


%



\bibitem{p2014}
Panik, M.J. (2014).
\textit{Growth Curve Modeling: Theory and Applications}, Wiley, New York.




\bibitem{p2009}
Pearl, J. (2009).
{\it Causality: Models, Reasoning, and Inference} (2nd Edition). Cambridge University Press, New York.

\bibitem{qz2014}
Qoyyimi, D.T. and Zitikis, R. (2014).
Measuring the lack of monotonicity in functions. {\it Mathematical Scientist} \textbf{39} (2), 107--117.




\bibitem{q1982}
Quiggin, J. (1982).
A theory of anticipated utility.
\textit{Journal of Economic Behavior and Organization} \textbf{3} (4), 323--343.


\bibitem{q1993}
Quiggin, J. (1993).
\textit{Generalized Expected Utility Theory: The Rank Dependent Model.}
Kluwer, Dordrecht.



\bibitem{r2013}
R Core Team (2013).
{\it R: A Language and Environment for Statistical Computing}. R Foundation for Statistical Computing, Vienna, Austria. ISBN 3-900051-07-0, URL http://www.R-project.org/.



\bibitem{rsw1995}
Ruppert, D., Sheather, S.J. and Wand, M.P. (1995).
An effective bandwidth selector for local least squares regression. {\it Journal of the American Statistical Association} \textbf{90} (432), 1257--1270.

\bibitem{rw1994}
Ruppert, D. and Wand, M.P. (1994).
Multivariate locally weighted least squares regression. {\it Annals of Statistics} \textbf{22} (3), 1346--1370.



\bibitem{r1983}
R\"{u}schendorf, L. (1983).
Solution of a statistical optimization problem by rearrangement methods.
\textit{Metrika} \textbf{30} (1), 55--61.



\bibitem{s1984}
Scarsini, M. (1984). On measures of concordance. \textit{Stochastica} \textbf{8} (3), 201--218.





\bibitem{sw1981}
Schweizer, B. and Wolff, E.F. (1981). On nonparametric measures of dependence for random variables. \textit{Annals of Statistics} \textbf{9} (4), 879--885.


\bibitem{sw1989}
Seber, G.A.F. and Wild, C.J. (1989). {\it Nonlinear Regression}. Wiley, New York.




\bibitem{ss2006}
Shaked, M. and Shanthikumar, J.G. (2007).
\textit{Stochastic Orders}. Springer, New York.







%



%


%


\bibitem{tt2010}
Thorndike, R.M. and Thorndike-Christ, T. (2010).
{\it Measurement and Evaluation in Psychology and Education} (8th Edition). Prentice Hall, Boston.


\bibitem{wj1995}
Wand, M.P. and Jones, M.C. (1995).
{\it Kernel Smoothing}. Chapman \& Hall, London.

\bibitem{wr2008}
Wand, M.P. and Ripley, B. (2014).
{\it KernSmooth: Functions for Kernel Smoothing}. R package version 2.23-13, URL http://CRAN.R-project.org/package=KernSmooth.




\bibitem{yj1997}
Yu, K. and Jones, M.C. (1997).
A comparison of local constant and local linear regression quantile estimator. {\it Computational Statistics \& Data Analysis} \textbf{25} (2), 159--166.








\end{thebibliography}
\end{document}